\documentclass[journal, 10pt]{IEEEtran}
\IEEEoverridecommandlockouts
\usepackage{tabularx}
\usepackage{cite}
\usepackage{amsmath,amssymb,amsfonts}
\usepackage{algorithmic}
\usepackage{graphicx}
\usepackage{textcomp}
\usepackage{xcolor,soul,framed}
\usepackage{comment}
\usepackage{multirow}
\usepackage{caption} 
\usepackage{amsmath}
\usepackage{booktabs}
\usepackage[linesnumbered,lined,boxed,commentsnumbered,ruled,longend, noend]{algorithm2e}
\SetKwInput{KwInput}{Input}               
\SetKwInput{KwOutput}{Output}              
\usepackage{array}
\usepackage{makecell}
\usepackage{amssymb}
\usepackage[dvipsnames]{xcolor}
\usepackage{hyperref}
\usepackage{caption}
\usepackage{setspace} 
\usepackage{enumitem}
\usepackage{fancyhdr}

\makeatletter
\def\ps@IEEEtitlepagestyle{%
  \def\@oddfoot{\mycopyrightnotice}%
  \def\@evenfoot{}%
}
\makeatother

\def\mycopyrightnotice{%
  \parbox{\columnwidth}{\footnotesize
  The manuscript has been published in IEEE Internet of Things Journal. Copyright~\copyright~2025 IEEE. Personal use of this material is permitted. However,\\
  permission to use this material for any other purposes must be obtained from the IEEE by sending a request to pubs-permissions@ieee.org.\hfill
  }
  \gdef\mycopyrightnotice{}
}

\begin{document}

\title{Towards Securing IIoT: An Innovative Privacy-Preserving Anomaly Detector Based on Federated Learning\\}

\author{Samira Kamali Poorazad, Chafika Benza\"{i}d, and Tarik Taleb%
    \thanks{Samira Kamali Poorazad and Chafika Benza\"{i}d are with the Centre for Wireless Communications, University of Oulu, Oulu, Finland (emails: samira.kamalipoorazad@oulu.fi, chafika.benzaid@oulu.fi).}%
    
    \thanks{Tarik Taleb is with the Department of Electrical Engineering and Information Technology, Ruhr University Bochum, Bochum, Germany (email: tarik.taleb@rub.de).}
}

\maketitle



\begin{abstract}

In the light of the growing connectivity and sensitivity of industrial data,  
 
cyberattacks and data breaches are becoming more common in the Industrial Internet of Things (IIoT). To cope with such threats, this study presents an anomaly detection system based on a novel Federated Learning (FL) framework. This system detects anomalies such as cyberattacks and protects industrial data privacy by processing data locally and training anomaly detection models on industrial agents without sharing raw data. The proposed FL framework incorporates two key components to enhance both privacy and efficiency. The first component is Homomorphic Encryption (HE), which is integrated into the framework to further protect sensitive data transmissions such as model parameters. HE enhances privacy in FL by preventing adversaries from inferring private industrial data through attacks, such as model inversion attacks. The second component is an innovative dynamic agent selection scheme, wherein a selection threshold is calculated based on agent delays and data size. The purpose of this new scheme is to mitigate the straggler effect and the communication bottleneck that occur in traditional FL architectures, such as synchronous and asynchronous architectures. It ensures that agents are not unfairly selected by the different delays resulting from heterogeneous data in IIoT environments, while simultaneously improving model performance and convergence speed.

The proposed framework exhibits superior performance over 

baseline approaches in terms of accuracy, precision, F1-scores, communication costs, convergence speeds, and fairness rate.

\end{abstract}


\section{Introduction}\label{sec:introduction}
\IEEEPARstart{M}{anufacturing} and industrial sectors use Internet of Things (IoT) technologies to automate processes and improve product quality through Industrial Internet of Things (IIoT) \cite{samira}. IIoT enhances productivity and scalability through intelligent interconnection and remote management \cite{samira}. However, due to the inherent broadcast nature of wireless communications, IIoT presents cybersecurity risks, such as command injection attacks \cite{attackiiot}. These attacks take the form of anomalies -- unexpected deviations from the system behavior -- which may indicate malicious activities or system malfunctions. The Stuxnet on Iran's nuclear power plant in 2010 is a high-profile example of these attacks \cite{stuxnet}. The risk of such attacks is especially pronounced in 
older industrial systems, which were not originally designed with security in mind.
\cite{ICSDetectAttack}. 
A robust anomaly detection system that continuously monitors and identifies potential attacks based on data flows in IIoT is crucial for mitigating potential security vulnerabilities. In this vein, many IIoT environments have recently adopted centralized machine learning-based anomaly detection methods  \cite{CommEfficient}. The use of these centralized approaches provides significant benefits in terms of improved model accuracy and ease of deployment \cite{smartbuildings}. However, they also introduce communication inefficiencies and raise potential privacy concerns, as 
large volumes of IIoT data must be transferred and processed at a central server \cite{smartbuildings}. In fact, requiring a central server to manage all data from IIoT agents increases the risks of data breaches and establishes a single point of failure \cite{DÏoT}.
Furthermore, IIoT data holders tend to avoid sharing sensitive information with third parties. Therefore, a practical, distributed anomaly detection system that protects data privacy in IIoT environments is imperative.

Federated Learning (FL), a form of distributed machine learning, offers a promising solution to meet the aforementioned need\cite{Zhao}. Through FL, industrial agents are able to work collaboratively to train global models by transmitting parameters and local models to a central server rather than sharing raw training data \cite{DÏoT}. Compared to centralized machine learning approaches, FL does not only safeguard privacy but also significantly reduces communication overhead, since sending only model parameters is much more efficient than sending the original training data. While FL significantly reduces privacy risk, adversaries can still compromise sensitive data by intercepting communication channels or compromising the aggregator server \cite{AIB5G}. Thus, FL remains vulnerable to critical threats such as model poisoning and model inference attacks \cite{AIB5G, Taleb2023AIB5G}. It is crucial that FL's privacy be further enhanced, particularly in the context of IIoT. Multiple secure FL schemes have been developed to address this need, including differential privacy (DP), Multiparty Computation (MPC), and homomorphic encryption (HE) \cite{FusionFL}.

The choice of communication mode --synchronous or asynchronous-- is as critical as addressing privacy concerns in FL frameworks. In heterogeneous IIoT environments, where devices operate at different speeds, due to varying computational resources and diverse data, the type of communication mode becomes even more significant. As an example, in FL with synchronous communication mode, all industrial agents must upload their local models to the server at the same time for aggregation \cite{FEDAVG}. This requirement forces the server 
to wait for the slowest device, commonly referred to as the \textit{straggler}, which delays the entire training process. This delay, known as the \textit{straggler effect}, makes Syncrounou Federated learning (SyncFL) unsuitable for real-time or heterogeneous IIoT applications, as it significantly slows down the convergence speed of the model and reduces the training efficiency of the anomaly detection model. The asynchronous FL (AsyncFL) \cite{asynchFL} was introduced to avoid straggler effects by performing global aggregation right after a local model has been received. Despite the frequent model transfer aggregations, AsyncFL approaches may be problematic for IIoT due to \textit{communication bottlenecks} caused by agents communicating with the aggregation server at different times, rather than in a coordinated manner. One solution for balancing SyncFL and AsyncFL is to use buffered-based solutions. Fed-Buff \cite{fedbuff} is an example of buffered FL, wherein local model updates are processed after a K-size buffer is filled at the server. A disadvantage of Fed-Buff is that it could favor agents with fast training speeds and results in low model accuracy. Recent research \cite{samiracape} proposes a Buffered FL (BFL) and an agent selection method based on the training time of agents in order to fill the gap in Fed-Buff. However, the authors in \cite{samiracape} considered only training time and used the agent selection method exclusively in the first training round, leading to two significant challenges.
First, considering that the computing capabilities of IIoT agents can fluctuate over time due to resource heterogeneity, 
selecting agents only during the first round of FL is neither dynamic nor efficient. Agents not selected in the initial round may have sufficient resources in later rounds and should be reconsidered for selection in subsequent rounds. Second, other factors, such as the communication delay, which is a variable factor between the agents and the central server should be considered. Therefore, focusing solely on training time is not sufficient. Consequently, there is a need for a comprehensive FL framework that addresses the highly heterogeneous IIoT environment by balancing the trade-offs between model convergence speed, accuracy, and varying agent speeds.

As a remedy to the above-mentioned challenges, a novel Dynamic HE-based FL (DyHFL) framework for the detection of IIoT anomalies is developed.  HE and an innovative Dynamic Agent Selection method are used to address three critical challenges, namely: privacy preservation, stragglers, and communication bottlenecks. HE is chosen for its ability to preserve 
both model accuracy and data privacy, whereas DP, despite its privacy benefits, tends to degrade model accuracy due to the introduced noise. 
MPC was excluded due to its high computational and communication overhead from constant data exchanges, causing delays that are unsuitable for real-time IIoT applications that require low-latency responses. In contrast, HE performs computations on encrypted data without continuous interaction, reducing overhead and making it more efficient for real-time IIoT environments. Furthermore, a novel \textbf{Dynamic Agent Selection} strategy is proposed to overcome the limitations of previous buffered FL methods, such as the static agent selection in BFL~\cite{samiracape} and the fixed buffer-based aggregation in FedBuff~\cite{fedbuff}.

 Unlike BFL~\cite{samiracape}, which only considers training time and performs agent selection once at initialization, DyHFL introduces a \textit{sliding window-based mechanism} that dynamically adjusts agent selection at every training round. This mechanism continuously evaluates agent performance based on three IIoT-relevant metrics: (1) \textit{training time}, which reflects computational capacity, (2) \textit{communication time}, which reflects the total round-trip latency, including model upload/download transmission with network latency, and (3) \textit{local data size}, which represents workload imbalance. This tri-metric evaluation enables DyHFL to adapt to heterogeneous IIoT devices with varying resources and conditions. In contrast to FedBuff ~\cite{fedbuff}, which fills buffers solely based on the arrival order of updates—often favoring faster devices—DyHFL employs a threshold-driven buffer update policy that allows aggregation to proceed without waiting for all agents. To operationalize this policy, DyHFL introduces two key functions—Weighted Average Metrics (WAM) ~\cite{samiracape} and Exponentially Weighted Moving Average (EWA). These functions combine three IIoT-relevant metrics (training time, communication time, and local data size) to compute a dynamic time threshold, which is then used to categorize agents as fast or slow. In this way, DyHFL ensures a fair representation of both groups in the aggregation buffer, rather than exhibiting bias toward faster agents as in FedBuff. This work mainly contributes to:

\begin{enumerate}
  \item Proposing a new method of anomaly detection that combines deep learning (DL) and HE to detect cyber threats in industrial cyber-physical systems (CPS). 
  \item Developing a novel FL framework that amalgamates the potential of synchronous and buffered FL approaches to effectively address straggler and communication issues while accounting for 
  the heterogeneous nature of data and resources in IIoT environments. 
  \item Designing an adaptive and dynamic agent selection method that continuously monitors training time, communication delay, and data size to fairly balance participation between fast and slow agents across training rounds.

    \item Evaluating the proposed DyHFL framework using three distinct industrial datasets to demonstrate its generalization ability, robustness, and superior performance compared to state-of-the-art FL baselines. 
\end{enumerate}

The remainder of this article is organized as follows. Section ~\ref{sec:sec2} discusses some related work in the literature. Section \ref{sec:method} introduces the system model, the attack model, and the proposed privacy-preserving FL method. Section \ref{secanal} discusses security analysis and communication complexity. Section \ref{eval} presents implementation details and discusses the evaluation results. Section \ref{disc} highlights limitations and potential future directions. This article concludes in Section \ref{conc}.

\section{Related Work}\label{sec:sec2}
This section  briefly reviews some studies focusing on FL-based anomaly detection systems for IIoT environments.

\subsection{FL-based Intrusion/Anomaly Detection}

\subsubsection{SyncFL approaches} In \cite{DÏoT}, the authors proposed a federated self-learning system for detecting malicious devices in IoT. Gated Recurrent Units (GRUs) are used in the system to classify data based on thresholds. Additionally, the self-learning mechanism is leveraged to enhance the detection performance of the global model as the IoT environment changes. For detecting energy efficiency anomalies in smart buildings, the authors of \cite{smartbuildings} proposed a FL approach based on Long Short-Term Memory (LSTM). In  \cite{ICSDetectAttack}, a Variational Autoencoder-LSTM model is used to detect anomalies in industrial control systems (ICS). The authors of \cite{CommEfficient}  devised a FL framework based on Convolutional neural network (CNN)-LSTM to reduce communication costs through gradient compression and local computations, wherein the Top-k algorithm is used to identify and send the "k" largest gradients for communication. The work in \cite{ensembleFL} presents a combination of  One-Class Support Vector Machine (OCSVM) and Isolation Forests (IF) to detect potentially malicious data points in IIoT. Stacked Autoencoders (SAE) are also used to extract features from data to help identify patterns and relationships.


The authors of \cite{ICT} propose a FL-based anomaly detection framework for smart grids, enabling smart meters to train local models without sharing data, ensuring privacy while maintaining comparable performance to centralized models. It evaluates seven ML classifiers, demonstrating low resource consumption, but lacks advanced privacy techniques (e.g., HE). In \cite{DNN} a FL framework for anomaly detection in IoT networks, integrating mutual information for feature selection and a deep neural network (DNN) for intrusion detection is proposed. It employs a mini-batch aggregation scheme to train models across distributed IoT devices, ensuring data privacy and high accuracy. However, the work in \cite{DNN} lacks robust privacy mechanisms against model inversion attacks, and favoring fast and well-connected devices due to its mini-batch aggregation approach.

Aside from the lack of privacy-preserving methods, the discussed synchronous methods are particularly susceptible to straggler effects when there are more agents involved. In real-time industrial domains, stragglers cause delays and slow convergence, thereby undermining the timely detection and mitigation of anomalies.

\subsubsection{AsyncFL approaches}

There are often impractical assumptions behind synchronized schemes. First of all, they require all nodes to transmit weights during every training round, resulting in significant network congestion. Second, these schemes require waiting for the slowest agents (i.e., stragglers), resulting in substantial delays. To address these limitations, AsyncFL approaches have been explored as a more efficient alternative. For example, in \cite{digital_twin_AFL}, the authors proposed an AsyncFL-based digital twin architecture for IIoT applications to minimize straggler effects. Results showed faster convergence and higher learning rates with the suggested model. Authors in \cite{edgeasync} introduced an AsyncFL scheme based on dynamically selected agents for heterogeneous IoT devices, which optimizes training by taking into account computing resources and network conditions. Compared to synchronous methods, the proposed AsyncFL is more efficient and accurate, and offers better scalability and robustness, especially in non-identical data scenarios. However, it introduces complexity and potential staleness.  In \cite{CSAFL}, the authors proposed a spectral clustering method based on the latency and direction of model updates to prevent model staleness. In non-independent and identically distributed datasets, the scheme also improved accuracy and convergence speed.  In \cite{SAFA}, through the use of a novel aggregation algorithm combined with a cache structure, the authors developed a Semi-Asynchronous Federated Averaging (SAFA) method to solve low round efficiency and poor convergence.  Authors in \cite{bilstm} suggested an AsyncFL model to detect low-rate Distributed Denial-of-Service (DDoS) attacks. This approach uses bidirectional LSTMs and an attention mechanism to improve detection accuracy by addressing missing data issues and ensuring the model learns from past and future data. The Proposed AsyncFL framework enables asynchronous updates across agents, reducing the impact of abnormal parameters and improving robustness. Study results indicate that the model outperforms state-of-the-art methods in terms of both accuracy and communication overhead, while potential implementation complexity was identified. Authors in \cite{DQN} introduce DEAFL-ID, a delay and energy-efficient AFL framework designed for intrusion detection in heterogeneous IIoT. DEAFL-ID leverages deep Q-learning (DQN) for optimal device selection to minimize training costs while preserving detection accuracy. However, the DEAFL-ID framework tends to prioritize faster devices since the selection criteria favor lower delay, higher energy efficiency, and better detection accuracy, which may introduce bias in the global model.

According to the investigated studies, asynchronous approaches are more accurate, scalable, and efficient in complex and non-uniform data environments. Nevertheless, despite their advantages, these methods are not without challenges, including staleness, increased complexity, and heightened communication costs. The server may slow down model convergence if it has to aggregate frequently. As a result, it is necessary to balance the advantages and disadvantages of AsyncFL approaches.

\subsubsection{Buffered FL approaches}

Buffer-based approaches have been introduced
to balance the limitations of SyncFL and AsyncFL frameworks \cite{fedbuff, samiracape}. However, these methods come with their own drawbacks.
In the approach proposed by \cite{fedbuff}, the authors overlooked the issue of fairness among agents, which can lead to the buffer being filled by faster agents, potentially sidelining slower, yet valuable contributors. On the other hand, while the work in \cite{samiracape} addressed fairness through an agent selection strategy, this solution has its limitations as it only applies during the first training round, lacking the flexibility to adapt dynamically throughout the learning process. Overall, there is a clear need for a more dynamic and efficient buffer-based solution to mitigate straggler by continuously adapting to the varying capabilities of agents, ensuring a fair and balanced contribution from all participants. In \cite{ASR-fed}, the authors introduce a buffered FL approach for anomaly detection in drone networks named  Agnostic  Straggler Resilient (ASR\_Fed), designed to handle straggling agents efficiently. The proposed method in \cite{ASR-fed} dynamically selects agents based on their accuracy and latency, ensuring that high-performing and responsive agents contribute early, while straggling agents are included later through a buffer-and-circumvent aggregation mechanism. Experimental results demonstrate that the proposed buffered methodology achieves higher accuracy and lower communication overhead compared to traditional FL algorithms. However, by initially involving only fast agents and delaying the participation of stragglers, ASR-Fed limits exposure to the full data distribution, thereby reducing update diversity. This constraint negatively affects model generalization, resulting in poorer performance and slower convergence. While stragglers are eventually included, their late contribution diminishes the impact of rare or critical data during the early learning stages.

\subsection{Privacy-enhancing FL} 

Though FL methodology offers significant privacy-preserving benefits, concerns about information leakage persist~\cite{Taleb2023AIB5G}. As outlined in \cite{dataleak}, sensitive participant information in distributed FL can be compromised, even when only model update data is shared. Publicly available gradients can sometimes be used to reconstruct original training data. To address these vulnerabilities, Privacy Enhancing Technologies (PETs) such as DP, MPC and HE have been introduced to enhance FL's effectiveness by securing shared model updates.

In \cite{SecPrivacyFL}, an IoT anomaly detection FL model incorporating DP is proposed to enhance privacy. To this end, the method employs a Generative Adversarial Network (GAN) to generate synthetic local model parameters and adds controlled noise to the raw local models. The study in \cite{HECovid-19} examines the use of FL and HE to protect sensitive data shared during the training of CNN models.  
The study in \cite{MK-CKKS} demonstrated that HE can provide comparable model accuracy to that of non-encrypted models while optimizing communication and computational efficiency. By employing an aggregated public key, the proposed method ensures the confidentiality of individual model updates without significantly affecting classification performance. Optimizing encrypted model updates is particularly useful for large-scale IoT deployments with limited resources. The Paillier Federated Multi-Layer Perceptron (PFMLP) proposed in \cite{PFMLP} integrates HE with FL to ensure the security of gradient data during transmission while maintaining accuracy close to traditional methods. Authors in \cite{VC} provided a secure FL framework incorporating HE and Verifiable Computing (VC) in order to ensure confidentiality and integrity in model training, especially in cross-silo settings with few reliable agents. However, the increased complexity and resource demands of this approach make it less suitable for resource-constrained environments, despite maintaining model accuracy and providing strong privacy and integrity guarantees. The approach in \cite{reliableFL} utilizes HE to enhance FL privacy, addressing common issues such as privacy breaches, communication overhead, and a lack of accountability. Meanwhile, the work in \cite{DeepFed} applied FL based on CNN and GRU in order to improve the accuracy of ICS intrusion detection by learning both spatial and temporal features of network traffic data, while employing HE as a means of improving privacy. In \cite{PEMPC2}, the authors presented a method called Partially Encrypted MPC for FL. The purpose of this approach is to reduce the high communication and computation costs associated with traditional MPC while preserving data privacy. To prevent sensitive information from being exposed during the aggregation of models, they selectively encrypt key parameter values or gradients. Despite improving efficiency, this approach may face complexities and challenges, particularly in dynamic environments with frequent model updates. Similarly, the work in \cite{DLG} presents a method that strengthens FL security against indirect gradient leakage using MPC. In this method, a two-round model decomposition process ensures that the central server receives only a modified version of the model, preventing data reconstruction. Although MPC improves model accuracy and strengthens privacy safeguards, it also increases communication costs and computational requirements. In \cite{CIoT}, the authors introduce a two-level privacy-preserving FL framework for attack detection in  Consumer Internet of Things (CIoT), integrating Partially HE for secure model aggregation. While achieving high detection accuracy with reduced false positives, the framework lacks any straggler mitigation strategy, making it vulnerable to delays or failures when some CIoT devices are slow or intermittently connected.

            \begin{table*}[ht]
                \centering
                \caption{Summary of Research Gaps in Federated Learning Studies}
                \resizebox{\textwidth}{!}{
                \begin{tabular}{>{\centering\arraybackslash}p{3.0cm} >{\centering\arraybackslash}p{1.5cm} >{\centering\arraybackslash}p{7.0cm} >{\centering\arraybackslash}p{1.3cm} >{\centering\arraybackslash}p{1.3cm} >{\centering\arraybackslash}p{1.3cm} >{\centering\arraybackslash}p{2.0cm}}
                    \toprule
                    \textbf{Papers} & \textbf{Approach} & \textbf{Straggler mitigation strategy} & \textbf{PETs} & \textbf{Straggler} & \textbf{Fairness} & \textbf{Communication Bottleneck} \\
                    \midrule
                     \cite{ICSDetectAttack}, \cite{CommEfficient},\cite{smartbuildings}, \cite{DÏoT},  \cite{ensembleFL},\cite{ICT}, \cite{DNN} & SyncFL & \(\times\) & \(\times\) & \(\times\) & N/A  & N/A \\

                    \cite{HECovid-19}, \cite{MK-CKKS}, \cite{PFMLP}, \cite{VC}, \cite{DeepFed}, \cite{CIoT} & SyncFL &  \(\times\) & HE & \(\times\) & N/A  & N/A \\
                    
                    \cite{PEMPC2},\cite{DLG} & SyncFL & \(\times\) & MPC & \(\times\)  & N/A   &   N/A  \\

                     &  &  &  &   &    &    \\

                    \cite{digital_twin_AFL}, \cite{CSAFL} & AsyncFL & Clustering agents by computing power and Training data availability, communication latency & \(\times\) &  \(\checkmark\) & N/A  & \(\times\) \\

                    &  &  &  &   &    &    \\

                    \cite{edgeasync} & AsyncFL & Greedy agent selection based on higher computing power and transmission delay & \(\times\) &  \(\checkmark\) & \(\times\)  & \(\checkmark\) \\
                    
                    &  &  &  &   &    &    \\
                                                           
                    \cite{SAFA} & AsyncFL & lag-tolerant model distribution agent selection& \(\times\) & \(\checkmark\)  & \(\times\) & \(\checkmark\) \\

                    &  &  &  &   &    &    \\

                    \cite{bilstm} & AsyncFL & Agent selection based on dataset size and model accuracy & \(\times\) & \(\checkmark\)  & \(\times\) & \(\checkmark\) \\
                    
                    &  &  &  &   &    &    \\
                    
                    \cite{DQN} & AsyncFL & Deep Q-Network based agent selection prioritizes high-accuracy, low-delay, and energy-efficient devices & \(\times\) & \(\checkmark\) & \(\times\) & \(\checkmark\) \\
                    
                    &  &  &  &   &    &    \\
                    
                    \cite{SecPrivacyFL} & AsyncFL & Basic asynchronous algorithm & DP & \(\checkmark\) & N/A  & \(\times\) \\

                   &  &  &  &   &    &    \\
                   
                    \cite{reliableFL} & AsyncFL & Basic asynchronous algorithm & HE & \(\checkmark\)  & N/A   &  \(\times\) \\
                    
                    &  &  &  &   &    &    \\
                    
                    \cite{fedbuff} & Buffer-FL & Buffering the first K received updates per round & DP & \(\checkmark\) & \(\times\)  & \(\checkmark\) \\

                    &  &  &  &   &    &    \\
                    
                    \cite{samiracape} & Buffer-FL &  Combination  of buffering and agent selection based on training time in the first round & HE & \(\checkmark\) & A/E &  \(\checkmark\) \\

                    &  &  &  &   &    &    \\

                    \cite{ASR-fed} & Buffer-FL & Buffering based on prioritizing fast agents for early updates, delaying stragglers &  \(\times\) &  \(\checkmark\) & A/E &  \(\checkmark\)\\

                    &  &  &  &   &    &    \\

                    \textbf{DyHFL} & Buffer-FL & Combination  of buffering, agent selection, and sliding window based on training time, communication time, and data size & \(\checkmark\) &  \(\checkmark\) & \(\checkmark\) & \(\checkmark\) \\

                    \bottomrule
                \end{tabular}
                }
                \label{tab:fl_gaps}
                
                \vspace{0.2cm} 
                \footnotesize
                Note: \(\checkmark\) – Addressed; \(\times\) – Not addressed; N/A – Not an issue ; A/E – Addressed but not enough.
            \end{table*}

Based on the investigated articles and our case which is IIoT environments, HE is preferred over DP and MPC due to its ability to protect model updates without compromising accuracy, which is vital for effective anomaly detection in IIoT environments. In fact, DP adds noise that can lower model accuracy. Meanwhile, MPC requires heavy communication and computation, making it unsuitable for quick-response scenarios in IIoT. In this work, the adopted HE scheme follows established approaches in the literature and is not intended to introduce a novel HE variant; rather, its role is to provide privacy preservation within DyHFL, enabling secure collaboration in IIoT without altering the core HE design.

Table \ref{tab:fl_gaps} provides a comprehensive review of the studied articles, highlighting their gaps and advantages. It outlines the FL approaches, PETs, straggler issues, fairness considerations among agents with different speeds, and communication bottlenecks present in each study. This analysis offers a clear comparison of different methods and their associated challenges, emphasizing areas where further research is needed. Based on a thorough review of existing literature and the identified gaps in Table \ref{tab:fl_gaps}, it has become apparent that studied methods lack a unified and efficient approach that provides a comprehensive solution to the straggler effect, communication bottlenecks, fairness, and privacy concerns. To fill this gap, this paper proposes a novel HE-based dynamic buffered FL framework for privacy-preserving anomaly detection in IIoT environments. 
The proposed framework aims 
to effectively balance the straggler effect and communication bottlenecks using a buffer and a novel dynamic agent selection strategy, while also enhancing privacy in environments with agents operating at different speeds. As a result, this approach, unlike existing solutions, optimally balances convergence speed, model performance, fairness, and communication costs.

\section{{DyHFL Framework}}\label{sec:method}
In this section, the system model, the attack model, and the proposed methodology are introduced. Table \ref{tab:notation} summarizes the notations used in this study.

\begin{table}[ht]
    \centering
    \caption{List of Notations}
    \begin{tabular}{ll}
        \hline
        \textbf{Symbol} & \textbf{Description} \\
        \hline
        $T$ & Number of Rounds \\
        $F$ & Communication Frequency (local epochs) \\
        $N$ & Number of Agents \\
        $N_{\text{sel}}$ & Selected agents \\
        $M$ & Model Size \\
        $m$ & Initial parameters\\
        $M_{agg}$ & Encrypted global parameters \\
        $SW$ & Sliding window size \\
        $B$ & Size of Buffer \\
        $\alpha$ and $\beta$ & Floating-point numbers whose sum equals 1\\
        $t\_st$ & Start training time \\
        $t\_et$ & End training time \\
        $t\_sc$ & Start communication time \\
        $t\_ec$ & End communication time \\
        ${\text{Data\_S}}$ & Local data size of the agent\\
        $T_{\text{Train\_T}}$  & Total Training time \\
        $T_{\text{Com\_T}}$ & Total communication time \\
        ${\text{Global\_MT}}$ & Global Metric\\
        
        $\text{ST\_Thrsh}$ & Short Term threshold \\
        $\text{LT\_Thrsh}$ & Long Term threshold \\
        $c$ & A constant number \\
        $ComCst$ & Communication Cost \\
        $P_{key}$ & Public key \\
        $S_{key}$ & Private key \\
        $\oplus$ & HE-based addition \\
        $\otimes$ & HE-based multiplication \\
        \hline
    \end{tabular}
    
    \label{tab:notation}
\end{table}

\subsection{System Model}

        \setlength{\textfloatsep}{0pt}
        \begin{figure}[htbp]
        \centerline{\includegraphics[width=3.7in]{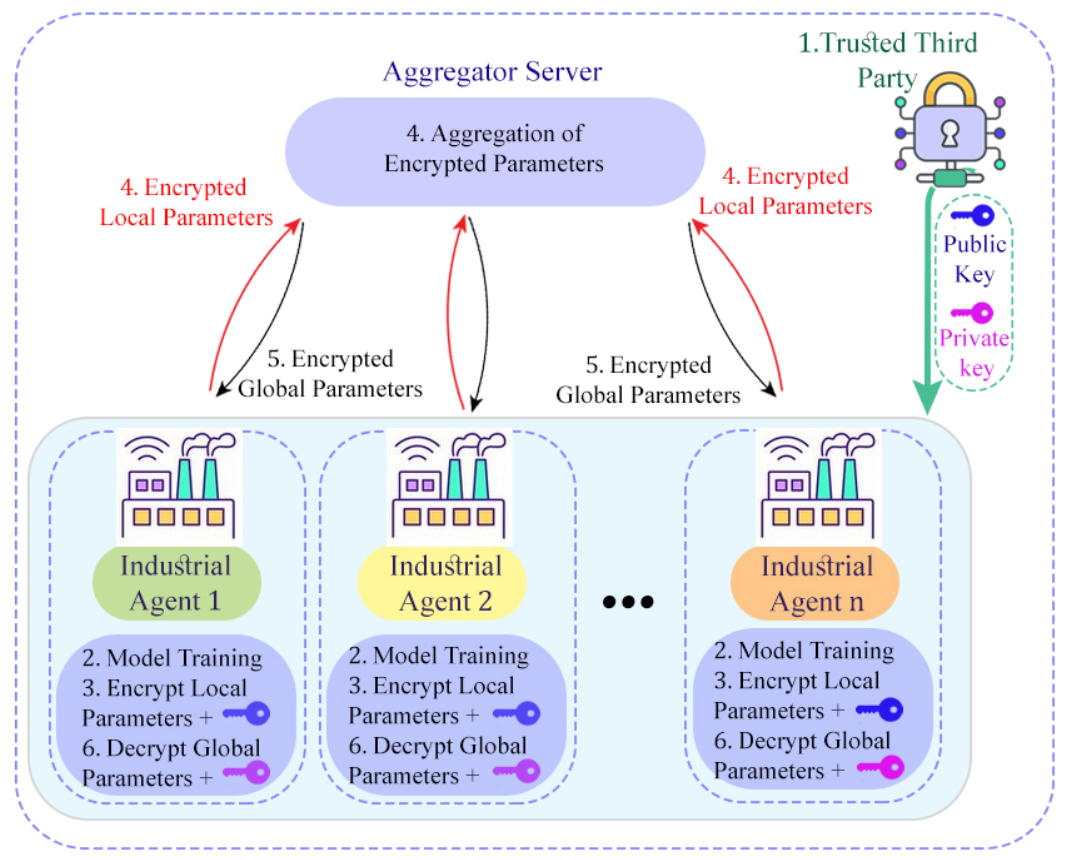}}
        \caption{High-level architecture of the DyHFL framework.}
        \label{figure:system_model}
        \vspace{0.5cm}
        \end{figure}

The system architecture in Fig. \ref{figure:system_model} is designed to ensure privacy while supporting FL across multiple industrial agents, each representing an owner of a CPS. It consists of the following steps:        
\begin{enumerate}
    \item \textbf{Key generation and Initialization:} At the beginning of the process, a trusted third party generates two cryptographic keys -- one public and one private -- for enabling secure communication between industrial agents and the FL aggregator server. The public key is used to encrypt data, while the private key is kept secret by each agent to decrypt data. The keys are then distributed to each industrial agent.

    \item \textbf{Local Model Training:} Industrial agents use their local CPS data to train DL models.
    \item \textbf{Encryption:} Once training is complete, the agent encrypts the model parameters (such as weights and biases) using the public key. In this way, sensitive model parameters are protected during transmission.

    \item \textbf{Secure Aggregation of Models:} Each agent's encrypted model parameters are sent to the server for aggregation. The aggregation server computes the encrypted global model by securely aggregating the encrypted parameters received from each participating agent.

    \item \textbf{Distribution of the Global Model:} The encrypted global model is sent back to all industrial agents once the aggregation process has been completed.
    \item \textbf{Decryption:} The global model is then decrypted by each agent using its private key. The local model of the agent is subsequently updated using the decrypted model.

\end{enumerate}

\subsection{Attack Model}

The threat model considered in this study addresses both cyber threats against industrial CPSs and adversarial attacks targeting the DyHFL framework. The study examines two main types of cyber threats facing the proposed model system:
\begin{enumerate}
 \item Cyber Threats Against Industrial CPSs: There are a number of threats to remote operations, including response injection attacks, command injection attacks, reconnaissance attacks, and denial of service (DoS) attacks. In response injection attacks, fake response messages are inserted into queries, whereas in command injection attacks, false commands are inserted into control systems. DoS attacks overload the target system, causing substantial disruptions to industrial processes, while reconnaissance attacks gather sensitive information about the CPSs.

\item Adversarial Threats Against the DyHFL Framework: This category focuses on two critical threats. In the first scenario, an "honest-but-curious" aggregation server may attempt to learn more than intended by inspecting the model parameters sent by agents. This could lead to the server inferring sensitive information about the agents' data or reverse-engineering the anomaly detection model, and compromising privacy in the FL process. The second concern involves external attacks or malicious eavesdroppers targeting communication links to intercept or alter the transmission of local parameters. This can expose sensitive data or manipulate updates, undermining the accuracy of the global model and threatening both the integrity and privacy of the DyHFL framework.
\end{enumerate}

\begin{algorithm}[]
\footnotesize
\caption{\scalebox{0.9}{\text{DyHFL-based Anomaly Detection}}}
\label{alg:detection_prevention}
\KwIn{$T$, $N$, $c$, $P_{key}$, $S_{key}$, $\alpha$ , $\beta$}
\KwOut{Desired performance (e.g., accuracy)}

$m \gets \text{Initial\_parameters()}$ \\
$M_{agg} \gets 0$ \ \# Encrypted global parameters \\
$SW \gets \frac{1}{c} \times T$ \\
\For{$P \gets 1$ \KwTo $SW$}{
    \For{each agent $k \gets 1$ \KwTo $N$}{
        $t\_st \gets \text{Start\_Time()}$ \\
        $Local_{\text{m}}^{(k)} \gets m.\text{train()}$ \\
        $m_{\text{enc}}^{(k)} \gets Encrypt(Local_{\text{m}}^{(k)}, P_{key})$ \\
        $t\_et \gets \text{End\_Time()}$\\
        $T_{\text{Train\_T}}^{(k)} \gets t\_et - t\_st$\\
        ${\text{Data\_S}}^{(k)} \gets Data\_Size()$\\
        $t\_sc \gets \text{Start\_Time()}$ \\
        send to server $(m_{\text{enc}}^{(k)})$
    }
    \# Encrypted aggregation on Server \\
    \For{each $j \in m_{\text{enc}}^{(k)}$}{
        $M_{\text{agg}} \gets (j \oplus M_{\text{agg}}) \otimes [N^{-1}]$
    } 
    send to agents $(M_{\text{agg}})$\\
    \For{each agent $k \gets 1$ \KwTo $N$}{
        Receiving $(M_{\text{agg}})$ by agents\\
        $t\_ec \gets \text{End\_Time()}$\\
        $T_{\text{Com\_T}}^{(k)} \gets t\_ec - t\_sc$\\
        $m_{\text{dec}}^{(k)} \gets Decrypt(M_{agg}, S_{key})$\\  
        
        send to server $(T_{\text{Com\_T}}^{(k)}, T_{\text{Train\_T}}^{(k)}, \text{Data\_S}^{(k)})$\\
    }

    \# Calculating Final Threshold on Server\\
    Normalize $(T_{\text{Com\_T}}^{(k)}, T_{\text{Train\_T}}^{(k)}, \text{Data\_S}^{(k)})$\\
    ${\text{Global\_MT}}^{(k)} \gets \alpha (T_{\text{Com\_T}}^{(k)} + T_{\text{Train\_T}}^{(k)}) + \beta ({\text{Data\_S}}^{(k)})$ \\
    $\scalebox{0.85}{$ST\_Thrsh^{(P)} \gets WAM\left({\text{Global}_{MT}^{(k)}}\right)_{k=1}^{N}$}$\\
    
}
${\text{LT\_Thrsh} \gets \scalebox{0.85}{$\left( \text{EWA}\left( \left( ST\_Thrsh^{(P)} \right)_{P=1}^{SW} \right) + \text{MAX}\left( \left( ST\_Thrsh^{(P)} \right)_{P=1}^{SW} \right) \right) / 2$}}$\\
\For{each $i \in {\text{Global\_MT}}^{(k)}$}{
    \If{$i \leq {\text{LT\_Thrsh}}$}{
        $[S_m] \gets i$
    }
}
\For{$S \gets SW + 1$ \KwTo $T$}{
    \For{each agent $k \in [S_m]$}{
        $Local_{\text{m}}^{(k)} \gets m.\text{train()}$ \\
        $m_{\text{enc}}^{(k)}\gets Encrypt(Local_{\text{m}}^{(k)}, P_{key})$
    }
 
}

\end{algorithm}
\normalsize

It is intended that by examining cyber threats, DyHFL will strengthen the security and resilience of industrial CPSs against a variety of malicious activities.

\normalsize

\setlength{\textfloatsep}{0pt}

\subsection {Proposed Methodology} \label{bb}
The proposed HE-based FL framework employs two key strategies to enable efficient training of DL anomaly detection models across multiple industrial CPS owners: a \textbf{synchronous communication mode} and a \textbf{dynamic buffer-based agent selection mechanism}.

On one hand, the framework adopts the \textbf{synchronous communication mode}to mitigate the communication bottlenecks commonly associated with AsyncFL. In AsyncFL, updated models must be transmitted individually to each industrial agent, leading to increased network overhead and inconsistencies in model convergence. By contrast, the synchronous approach ensures that model updates are aggregated and distributed in a coordinated manner, significantly reducing network load and promoting more stable global model updates.

On the other hand, the framework introduces a \textbf{buffer-based agent selection mechanism } to reduce delays caused by slow agents, commonly known as the straggler effect in SyncFL. In synchronous FL, the server must wait for all selected agents to return their local model updates before aggregation. This can significantly hinder training progress. Proposed mechanism resolves this by introducing a buffer and a time threshold: the aggregation server collects model updates into a buffer and begins aggregation once the specified threshold time has elapsed, regardless of whether all agents have responded. This mechanism not only minimizes training delays but also mitigates bias introduced by uneven agent participation, leading to more balanced model performance. The dynamic buffer-based selection mechanism implemented by dividing the total number of training rounds \textit{T} into two distinct phases: \textbf{Preliminary Rounds}, used to estimate the time threshold, and \textbf{Subsequent Rounds}, where agent selection is guided by this threshold.

\begin{enumerate}

\item \textbf{Phase One: Preliminary Rounds (Rounds 1 to SW)} 


The objective of the preliminary rounds is to estimate a time threshold that will guide agent selection in subsequent training rounds. This is achieved using a sliding window mechanism, designed to monitor agent-specific performance metrics over time and ensure fairness and adaptability. Each agent records the following performance metrics that will be used in computing the time threshold:
\begin{itemize}
    \item \textbf{Data size:} the volume of local data available for training.
    \item \textbf{Training Time:} the duration taken to train the local model.
    \item \textbf{Communication Time:} the time required to transmit the encrypted model to the server and receive the updated global model.
\end{itemize}

Below is a  detailed Step-by-Step Workflow:

\begin{enumerate}
                        \item \textbf{sliding window size calculation:} In the first step, \textit{the sliding window size (SW)} is set equal to the number of \textit{preliminary rounds (P)}, which is computed as:

                                    {\footnotesize
                                    \begin{equation}
                                      SW= P = \frac{T}{c}
                                    \end{equation}
                                    }
                            
                            where \textit{T} is the total number of training rounds, and \textit{c} is a constant that controls the proportion of rounds allocated to the threshold estimation phase (line 3 of Algorithm~\ref{alg:detection_prevention}). 
                            
                        \item \textbf{Model Training:} Once the public and private keys are generated, agents obtain parameter values from the aggregator server. The agents then train their DL models using their respective data in order to detect anomalies (line 7 of Algorithm \ref{alg:detection_prevention}).
                    
                        \item  \textbf{Monitoring Metrics:} During the training process, the amount of time that each agent spends on training along with the size of data are measured (lines 10 and 11 of Algorithm \ref{alg:detection_prevention}).

                        \item  \textbf{Model Encryption and Transmission:}
                        Once the local model is trained, the parameters are encrypted using the public key (line 8 of Algorithm \ref{alg:detection_prevention}). Following training and encryption, the encrypted parameters are sent to the aggregator server (line 13 of Algorithm \ref{alg:detection_prevention}). This is the point at which the calculation of communication time begins until the agents receive the updated encrypted global model (line 12 of Algorithm \ref{alg:detection_prevention}).   
                    
                        \item \textbf{Aggregation:} Aggregator server collects encrypted local parameters from agents and compute encrypted global parameters (line 16 of Algorithm \ref{alg:detection_prevention}). These encrypted global parameters are then sent back to the agents (line 17 of Algorithm \ref{alg:detection_prevention}).

                        \item \textbf{Updating and Decryption:} Agents receive the encrypted global parameters, which marks the end of the calculation of communication times (lines 19 and 20 of Algorithm \ref{alg:detection_prevention}). Afterward, using private keys, the agents decrypt the encrypted global parameters and update the local model parameters accordingly (line 22 of Algorithm \ref{alg:detection_prevention}).

                        
                        \item \textbf{Short-Term Threshold (ST\_Thrsh):} This step involves calculating the \textit{Short-Term Threshold (ST\_Thrsh)} using agent-specific performance metrics. The purpose of \textit{ST\_Thrsh} is to ensure fair participation by adjusting selection bias between fast and slow agents. To calculate the \textit{ST\_Thrsh}:

                        \begin{itemize}
                        \item Each agent’s performance metrics values (data size, training time, and communication time) are sent to the server  (as described in line 23 of Algorithm \ref{alg:detection_prevention}).   

                        \item The performance metrics values are normalized by the server using the \textit{Min-Max scaler}, which rescales the values to the range of [0, 1] (line 25 of Algorithm \ref{alg:detection_prevention}). The Min-Max scaler is defined as follows:

                            {\footnotesize
                            \begin{equation}
                            X_{\text{scaled}} = \frac{X - X_{\min}}{X_{\max} - X_{\min}}
                            \end{equation}
                            }
                    
                        where \( X \) represents the original value, \( X_{\min} \) is the minimum value within its respective list (training time, communication time, or data size), and \( X_{\max} \) is the maximum value within the same list.
                        
                        Normalization is necessary to balance the impact of different parameters, as they exist on different scales. Without normalization, larger numerical values (e.g., data size) could dominate the global metrics, leading to biased agent selection. 
                        
                        \item The normalized metrics are then summed to compute the agent’s global metric \textit{\textbf{(Global\_MT)}} (line 26 of Algorithm \ref{alg:detection_prevention}), as follows:
                        
                        {\footnotesize
                        \begin{equation}
                        \text{Global\_MT} = \alpha \cdot (T_{\text{Train\_T}} + T_{\text{Com\_T}}) + \beta \cdot \text{Data\_S}
                        \end{equation}
                        }
                        
                        where $\alpha$ and $\beta$ are floating-point numbers whose sum equals one $(\alpha + \beta = 1 )$, serving as weighting factors and normalization coefficients to balance the contributions of different components in the computation of the \textit{Global\_MT}. Their primary purpose is to ensure a controlled trade-off between the communication and training time components ($T_{\text{Train}\_T + T_{\text{Com}\_T}}$) and the dataset size component ($\textit{Data\_S}$). This prevents any single component from disproportionately influencing the \textit{Global\_MT}. By adjusting $\alpha$ and $\beta$, the model can emphasize either computation time or dataset size depending on the specific requirements of the system, ensuring an adaptive and efficient evaluation of the agent's performance.
                        
                        \item The computed $\textit{Global\_MT}$ values are fed into the \textit{WAM} function to derive the \textit{ST\_Thrsh}. \textit{WAM} assigns lower weights to faster agents and higher weights to slower ones, thereby promoting fair participation and preventing consistent exclusion from training. These weights are directly determined by each agent’s normalized global metrics (line 27 of Algorithm \ref{alg:detection_prevention} and Algorithm \ref{alg:weighted_time}).

                        \end{itemize}
                        
                        \item \textbf{Long-Term Threshold (LT\_Thrsh):} In the final step of the preliminary phase, the \textit{Long-Term Threshold (LT\_Thrsh)}—used as the time threshold for buffer-based agent selection—is computed using an \textit{EWA} applied to the sequence of \textit{ST\_Thrsh} values (line 28 of Algorithm~\ref{alg:detection_prevention} and Algorithm~\ref{alg:Exponential_time}). \textit{EWA} emphasizes recent \textit{ST\_Thrsh} values have a greater impact on the  \textit{LT\_Thrsh}, enabling the threshold to adapt dynamically while smoothing out transient fluctuations. The resulting \textit{LT\_Thrsh} serves as the final decision boundary for agent selection in subsequent training rounds.

                        \begin{figure}[htbp]
                            \centerline{\includegraphics[width=3.83in]{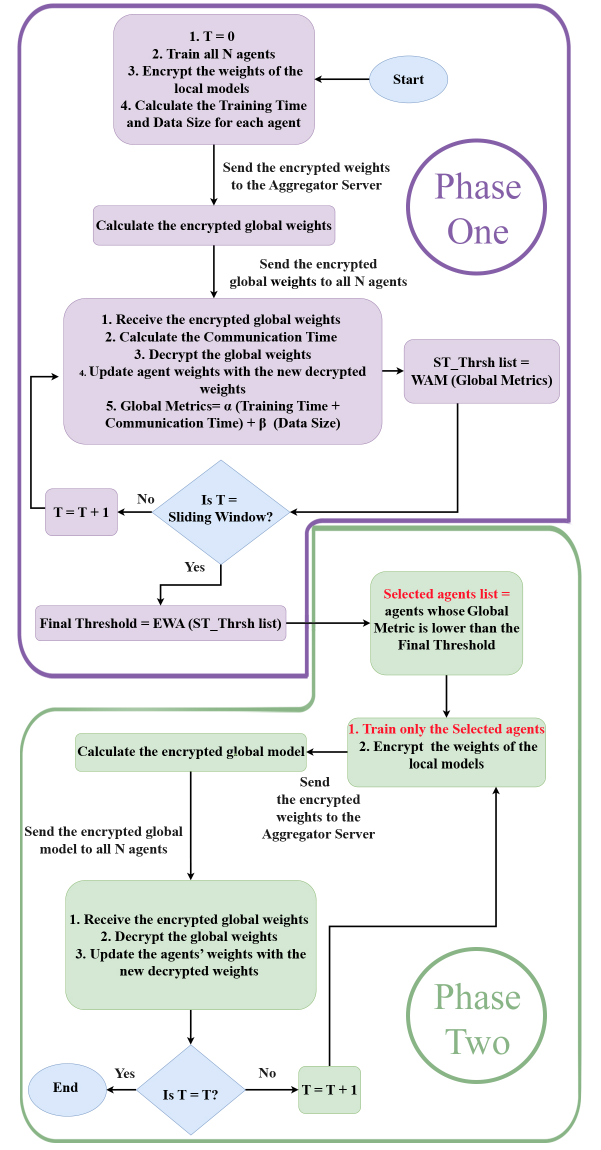}}
                            \vspace{-0.1cm}
                            \caption{ DyHFL framework Flowchart.}
                            \label{figure:acccomparison}
                            \vspace{-0.1cm}
                        \end{figure}
                    
                    \end{enumerate}

\item \textbf{Phase Two: Subsequent Rounds (Rounds SW+1 to T)} 

In this phase, the calculated time threshold (LT\_Thrsh) from preliminary round is employed by the aggregation server to govern agent selection and aggregation behavior (lines 29 to 35 of Algorithm~\ref{alg:detection_prevention}). This dynamic approach ensures that training progresses efficiently without being hindered by stragglers, communication bottlenecks, and still preserving fairness and representation across the agent population. 

Algorithms \ref{alg:detection_prevention}, \ref{alg:weighted_time}, \ref{alg:Exponential_time} summarize the detailed training procedure. 
\end{enumerate}

\begin{algorithm}[]
    \footnotesize
    \caption{\scalebox{0.9}{Weighted\_Average\_Metrics (WAM)}} 
    \label{alg:weighted_time}
    
    \KwInput{\scalebox{0.9}{Global metrics for each Agent $\gets {\text{Global\_MT}}^{1}, \ldots, {\text{Global\_MT}}^{N}$}}
    \KwOutput{Short-term threshold for each preliminary round}
    
    Sort ${{Global\_MT}}^{1},{{Global\_MT}}^{2},..., {{Global\_MT}}^{N}$ \ in\ descending\ order
    
    \For{each $i \gets 1$ to $n$}{
        $\text weight^{(i)}\gets \frac{1}{{\text{${Global\_MT}$}}^{i}}$\\
    }

    \# Reverse weight = $[W_1, W_2, \ldots, W_n]$ \\
    $\text{weight} \gets [W_N, W_{N-1}, \ldots, W_1]$
    
    \normalsize ${\text{W\_Avg}} \gets \frac{\sum_{j=1}^{n}{{\text{Global\_MT}}^{(j)} \cdot \text{weight}^{(j)}}}{\sum_{j=1}^{n}{\text{weight}^{(j)}}}$
    
\end{algorithm}

Steps \textit{\textbf{b}} through \textit{\textbf{g}} of phase one are repeated until the number of preliminary rounds (P) reaches T/c,  while the subsequent rounds of phase two continue until convergence. For instance, if T=100 and c=5, then the number of preliminary rounds will be P=20, which also defines the size of the sliding window (SW). The remaining 80 rounds constitute the subsequent phase of training. In this setup, the first 20 rounds are allocated to the preliminary phase, during which the system monitors agent behavior and calculates the time threshold. The following 80 rounds proceed with selected agents based on the computed threshold, continuing until convergence.

                    \begin{algorithm}[]
                    \footnotesize
                    \caption{\scalebox{0.9}{\text{Exponential\_Weighted\_Average (EWA)}}
                    }
                    \label{alg:Exponential_time}
                    \KwInput{\scalebox{0.90}{Short-term thresholds $\gets ST\_Thrsh^{1}, ST\_Thrsh^{2}, \ldots, ST\_Thrsh^{P}$}}
                    
                    \KwOutput{Final Threshold for short-term thresholds}
                    
                    {
                       
                        \For{each $i\gets1$ to $n$}{
                            
                            $\text{E\_Weight}^{(i)} \gets \frac{i * (i+1)}{{\text{$2$}}}$
                    
                        }   
                           $
                            \text{Thereshold} \gets \frac{\sum_{j=1}^{n} \text{E\_weight}^{(j)} * \text ST\_Thrsh^{(j)}}{\sum_{j=1}^{n} \text{E\_weight}^{(j)}}
                            $

                    }
                    
                    \end{algorithm}

\section {Security and Communication Complexity Analysis} \label{secanal}
This section provides a security and communication complexity analysis of the DyHFL framework. It demonstrates how HE safeguards model updates from both honest-but-curious servers and potential external eavesdroppers, while also providing a detailed comparison of the communication cost and computational complexity across different FL methods.

\subsection{\textbf{Protection Against an Honest-But-Curious Server}}

In DyHFL, the central server is assumed to be \textit{honest-but-curious}—it correctly follows the protocol but tries to learn private information from the data it receives. To prevent this, each agent encrypts its local model updates using a public key before sending them to the server.

\subsubsection*{\textbf{How It Works}}

Let \(\Delta w_i\) be the model update from agent \(i\), and \(P_{\text{key}}(\cdot)\) be the encryption function using a public key. Each agent sends:

{\footnotesize
\begin{equation}
P_{\text{key}}(\Delta w_i)
\end{equation}
}

The server receives encrypted updates from all \(N\) agents:

{\footnotesize
\begin{equation}
P_{\text{key}}(\Delta w_1),\; P_{\text{key}}(\Delta w_2),\; \dots,\; P_{\text{key}}(\Delta w_N)
\end{equation}
}

It then performs \textbf{homomorphic aggregation}, which gives an encrypted sum of the model updates:

{\footnotesize
\begin{equation}
P_{\text{key}}\!\left(\Delta W\right) =
P_{\text{key}}\!\left(\sum_{i=1}^{N} \Delta w_i\right)
\end{equation}
}

The server cannot decrypt the encrypted model updates, as it does not hold the private key. The encrypted aggregated model \( P_{\text{key}}(\Delta W) \) is then sent back to the agents, who can decrypt it using the private key \( S_{\text{key}} \), which is exclusively held by agents:

{\footnotesize
\begin{equation}
\Delta W = S_{\text{key}}\!\left(P_{\text{key}}(\Delta W)\right)
\end{equation}
}

Due to the \textbf{semantic security} of HE~\cite{semantic}, which guarantees that without the private key the server cannot distinguish between any two encrypted values, the server cannot infer any information about the individual encrypted model updates, even though it performs the aggregation. This ensures the privacy of individual agents, even in the presence of an honest-but-curious server.

\subsection{\textbf{Security Against Eavesdropping During Transmission}}

HE also protects model updates during communication. This prevents external attackers from learning private information by intercepting data.

\subsubsection*{\textbf{How It Works}}

Consider an external attacker \( \mathcal{A} \) attempting to eavesdrop on the communication between agents and the server by intercepting model updates during transmission. Suppose \( \mathcal{A} \) captures a ciphertext \( P_{\text{key}}(\Delta w) \). Even if provided with two candidate plaintext model updates, \( \Delta w_0 \) and \( \Delta w_1 \), and their corresponding ciphertexts \( P_{\text{key}}(\Delta w_0) \) and \( P_{\text{key}}(\Delta w_1) \), the adversary cannot determine which ciphertext corresponds to which plaintext with non-negligible probability.

HE is \textbf{probabilistic}, meaning each encryption of the same plaintext gives a different ciphertext. Its security relies on the decisional composite residuosity assumption, a hard problem in number theory. This assumption guarantees that, without the private key, an adversary cannot distinguish between the ciphertexts of two different plaintexts.

\subsection{\textbf{Communication Cost Complexity}}

Communication cost refers to the total size of messages exchanged, specifically models transmitted between agents and the aggregator server, measured in megabytes (MB). The communication cost (\textbf{ComCst}) for various FL methods is calculated as follows:

{\footnotesize
\begin{equation}
\text{FedBuff ComCst} = M \times T \times B
\end{equation}

\begin{equation}
\text{SyncFL ComCst} = M \times T \times N
\end{equation}

\begin{equation}
\text{BFL ComCst} = (M \times 1 \times N) + (M \times (T - 1) \times N_{\text{sel}})
\end{equation}

\begin{equation}
\text{ASR\_Fed ComCst} = M \times (T \times \text{Buffs} + (T - \text{CirT}) \times \text{Cirs})
\end{equation}

\begin{equation}
\text{DyHFL ComCst} = (M \times SW \times N) + (M \times (T - SW) \times N_{\text{sel}})
\end{equation}

\begin{equation}
\text{AsynFL ComCst} = M \times T \times N \times F
\end{equation}
}

Where:
\begin{itemize}
    \item \(M\): Model size
    \item \(T\): Total rounds
    \item \(N\): Number of agents
    \item \(B\): Buffer size
    \item \(N_{\text{sel}}\): Number of selected agents
    \item \(SW\): Sliding window size
    \item \text{Buffs}: buffer agent size in ASR\_Fed
    \item \text{Cirs}: circumvent agent size 
    \item \text{CirT}: circumvent threshold size
    \item \(F\): Communication frequency in AsyncFL which refers to the average number of communication events—both uploads and downloads—between each agent and the server per training round.

\end{itemize}

Table \ref{table:complexity} compares the time complexity $\mathcal{O}$ (Big O) and the best-case performance $\Omega$ (Big Omega) for various FL algorithms. Regarding \textbf{DyHFL} and  \textbf{BFL}, the best-case complexity occurs when the number of selected agents \( k \) is minimal, resulting in a complexity of $\Omega(\text{T} \times k)$. 

\begin{table}[h!]
    \caption{Computational Complexity of FL Baselines }
    \label{table:complexity}
    \scriptsize
    \centering
    \begin{tabular}{|l|c|c|}
        \hline
        \textbf{Algorithm} & \textbf{Big O} & \textbf{Big Omega} \\
        \hline
        DyHFL & \(O(T \times n)\) & \(\Omega(T \times k)\) \\
        BFL & \(O(T \times n)\) & \(\Omega(T \times k)\) \\
        FedBuff & \(O(T \times n)\) & \(\Omega(T \times B)\) \\
        SyncFL & \(O(T \times n)\) & \(\Omega(T \times n)\) \\
        AsyncFL & \(O(T \times n \times F)\) & \(\Omega(T \times n)\) \\
        ASR\_Fed & \(O(T \times \text{Cirs})\) & \(\Omega(T \times \text{Buffs})\) \\
        \hline
    \end{tabular}
\end{table}

In \textbf{Fedbuff}, the complexity is minimized when the buffer size is minimal, giving a complexity of $\Omega(\text{T} \times B)$. For \textbf{ASR\_Fed}, the complexity is minimized when the buffer agent size (Buffs) is maximized, resulting in a lower bound complexity of $\Omega(T \times Buffs)$. Conversely, the complexity is maximized when the circumvent agent size (Cirs) is maximized, leading to an upper bound complexity of $\mathcal{O}(T \times Cirs)$. The \textbf{SyncFL} method's complexity remains relatively stable across different scenarios, with a best-case of $\Omega(\text{T} \times n)$. Lastly, in \textbf{AsyncFL}, the best case occurs when the number of communication frequencies is minimized, typically to one, leading to a complexity of $\Omega(\text{T} \times n)$.\\

\section{Experiment Setting and Evaluation}\label{eval}
\vspace{-0.05cm}
\subsection{Experimental Settings}

The proposed FL framework was evaluated using three datasets with different characteristics in terms of number of classes, the size of the data, and the distribution of the samples, namely Gas\_Pipeline dataset \cite{gaspip}, the WUSTL\_IIoT dataset \cite{wsul-IIoT}, and the Edge\_IIoT  dataset \cite{edgeiiot}, as described in [15] to achieve a more comprehensive and reliable evaluation. The Gas\_Pipeline dataset contains industrial data with specific features, while the WUSTL\_IIoT and Edge\_IIoT datasets covers IoT-related data. Using these three diverse datasets allows us to demonstrate that DyHFL is not dependent on a specific type of data and can maintain strong, and reliable performance across different data volumes, distributions, and structural complexities. This approach effectively demonstrates the generalization of the proposed method and prevents concerns about dataset-specific overfitting. PyTorch was used to implement the proposed FL framework, while Paillier was used to implement HE. For demonstrating the effectiveness of the proposed model in scenarios involving heterogeneous data, non-identical datasets are used. The non-identical data distribution allows us to identify which agents are fast and which ones are slow, allowing us to simulate performance differences more accurately. Accordingly, FedArtML is used for the generation of non-identical datasets (Dirichlet and no-label-skew). \textbf{\textit{Dirichlet Method} }utilizes the Dirichlet distribution to create unequal distributions, resulting in agents receiving different proportions of data and varying percentages of labels. \textbf{\textit{No-Label-Skew Method} }ensures that each agent receives data with the same labels, thereby the distribution of labels remains consistent. Due to this, agents with larger data may have longer global times, known as slow agents, whereas agents with smaller data may have shorter global times, known as fast agents.

A pre-processing operation was conducted in order to clean the input data and improve its accuracy. Gas\_Pipeline dataset features with only one value are removed, reducing the number of features from 27 to 18. Datasets are split into three parts: $80\%$ for training, $10\%$ for validation, and $10\%$ for testing. Min-Max scaling is used to normalize the dataset's features.

The multilayer perceptron (MLP)-based IDS is implemented after performing data pre-processing. MLP model with three fully connected layers was used to analyze the Gas\_Pipeline, WUSTL\_IIoT, and Edge\_IIoT datasets. In the case of the Gas\_Pipeline dataset, the model's first layer transforms the 18-dimensional input into a 54-dimensional space. The second layer reduces this to a 20-dimensional space, and the final layer maps it to eight classes. In the case of WUSTL\_IIoT, the first layer transforms the 41-dimensional input into a 9-dimensional space, the second layer uses the same 9-dimensional space, and the final layer maps it to five classes. Regarding the Edge\_IIoT dataset, the model's first layer transforms the 15-dimensional input into a 64-dimensional space. The second layer reduces this to a 32-dimensional space, and the final layer maps it to fifteen classes. The model structure was determined based on the nature of the datasets (dimensionality and class distribution), experiments with different layer sizes, and the structure presented in \cite{samiracape}. 

Stochastic Gradient Descent (SGD) was chosen as the optimization algorithm, with a batch size of 64 for the Gas\_Pipeline dataset and 1000 for the WUSTL\_IIoT and Edge\_IIoT datasets. The learning rate was set to 0.01, and the momentum was set to 0.8. Based on best practices for the datasets, these settings were selected to ensure stable and efficient convergence. To optimize further, future work could explore different momentum settings or adaptive learning rates.

The proposed DyHFL framework uses a semi-synchronous buffer-based strategy. Therefore, to highlight its advantages, it is compared with five FL methods: SyncFL\cite{FEDAVG} and AsyncFL\cite{asynchFL} as traditional baselines, and with FedBuff\cite{fedbuff}, ASR\_Fed\cite{ASR-fed}, and BFL\cite{samiracape} as state-of-the-art buffered or selective aggregation methods. This comparison illustrates improvements over both fully synchronous/asynchronous schemes and existing buffered approaches. Detailed descriptions of \textbf{SyncFL, AsyncFL, and BFL} implementations can be found in the referenced sources. A buffer size equal to 75\% of total agents is recommended for \textbf{FedBuff} implementation, based on the information provided in Table \ref{table:fedbuffsizewustl}. The details of this choice are described in the Evaluation part of Section  \ref{eval}.

The \textbf{ASR\_Fed }implementation classifies agents into buffer agents and circumvent agents. Buffer agents include those whose accuracy and training time exceed a predefined threshold, while circumvent agents consist of those that do not meet this threshold. The accuracy and training time threshold is dynamically determined based on the average accuracy and training time of all participating agents. Agents categorized as buffer agents are considered fast agents, whereas those in the circumvent list are classified as slow agents. During the initial aggregation rounds, only updates from fast agents are incorporated. Slow agents participate in the aggregation process only when the total number of rounds reaches a predefined value, referred to as the circumvent threshold. This threshold is computed using the following formula based on reference \cite{ASR-fed}:

{\footnotesize
\begin{equation}
\text{Circumvent Threshold} = \text{int} \left( \left\lceil \frac{\sum \text{(circumvent agent delays)}}{\text{buffer agent count}} \right\rceil \right)
\end{equation}
}

Once the total number of rounds equals the circumvent threshold, slow agents also contribute to the aggregation process, ensuring a balanced and adaptive participation mechanism.

Four key metrics are evaluated in order to highlight the advantages of the proposed approach:  \textbf{1. Agent Selection Fairness, 2. Convergence Speed, 3. Communication Costs, and 4. Model Performance}.

\begin{enumerate}

  \item As mentioned earlier, one of the goals of the proposed method is to balance the participation between fast and slow (straggler) agents. \textbf{Agent selection fairness} is defined as assessing the fairness of the proposed method in terms of the participation of both fast and straggler agents across the entire agent population. To ensure that the proposed approach guarantees fair participation and performance for both fast and straggler agents, we consider two metrics: Straggler Rate Selection (SRS) and Fast Rate Selection (FRS). The formulas for these metrics are as follows: \\
   \begin{center}
   
   {\footnotesize
   \begin{equation}
    \text{SRS} = \frac{\text{Number of Selected Straggler Agents}}{\text{All of Straggler Agents}}
    \end{equation}
    
   \begin{equation}
    \text{FRS} = \frac{\text{Number of Selected Fast Agents}}{\text{All of Fast Agents}}
    \end{equation}}

    \end{center}
SRS measures the proportion of selected straggler agents compared to the total number of straggler agents available. The FRS represents the proportion of selected fast agents relative to all fast agents in the system.

\item \textbf{Convergence speed} is a measure of the number of rounds required to achieve a specified level of accuracy (i.e., target accuracy). When an approach requires fewer rounds to achieve the target accuracy, this indicates a faster rate of convergence, and vice versa. The purpose of this metric is to assess the robustness of FL methods against stragglers as well as their effectiveness in communication efficiency. This study compares different methods using 20 agents as a baseline, and a \textbf{target accuracy of 94.6 \% for the Gas\_Pipeline dataset, 99.8 \% for the WUSTL\_IIoT dataset and 98.5 \% for the Edge\_IIoT dataset}. In both cases, the target accuracy is set slightly above the baseline accuracy that the non-federated method achieves with centralized data, ensuring that the FL method performs as well, if not better, with distributed data. 

\item \textbf{Communication Cost}, as defined in Section~\ref{secanal}.C, denotes the total size of data exchanged—specifically the model parameters transmitted between agents and the aggregator server—measured in megabytes (MB).

\item \textbf{Model Performance} is evaluated using three key classification metrics: Accuracy, Precision, and the F\textsubscript{1} Score. These metrics are defined as follows:

\begin{itemize}
    \item \textbf{Accuracy} reflects the overall correctness of the model, calculated as the proportion of correctly predicted instances (both positive and negative) out of all predictions:
    {\footnotesize
    \begin{equation}
    \text{Accuracy} = \frac{\text{True Positives} + \text{True Negatives}}{\text{Total Instances}}
    \end{equation}
    }

    \item \textbf{Precision} quantifies the proportion of correctly predicted positive instances out of all predicted positives. It measures the reliability of the model’s positive predictions:
    {\footnotesize
    \begin{equation}
    \text{Precision} = \frac{\text{True Positives}}{\text{True Positives} + \text{False Positives}}
    \end{equation}
    }

    \item \textbf{F\textsubscript{1} Score} provides a single performance metric by taking the harmonic mean of Precision and Recall. 
    Recall measures the model’s ability to identify all actual positive instances and is defined as:
    {\footnotesize
    \begin{equation}
    \text{Recall} = \frac{\text{True Positives}}{\text{True Positives} + \text{False Negatives}}
    \end{equation}
    }
    The F\textsubscript{1} Score is then calculated as:
    {\footnotesize
    \begin{equation}
    \text{F}_1 \text{ Score} = 2 \cdot \frac{\text{Precision} \cdot \text{Recall}}{\text{Precision} + \text{Recall}}
    \end{equation}
    }

\end{itemize}

The F\textsubscript{1} Score is particularly useful in imbalanced classification problems, where it provides a more informative evaluation than Accuracy by considering both false positives and false negatives.

\end{enumerate}

\subsection{Evaluation}\label{jijiji}

The results were averaged from four independent runs for the metrics described earlier (i.e., Agent Selection Fairness, Convergence Speed, Communication Costs, and Model Performance metrics).

The first step in the evaluation process involves determining the optimal sliding window size and the corresponding values of $\alpha$ and $\beta$, which are essential for computing the DyHFL method results. As shown in Table  \ref{table:slidegas},  sliding window sizes of $1/10$, $1/5$, and $1/2$ of the total training rounds were evaluated alongside various combinations of $\alpha$ and $\beta$ values, specifically (0.3, 0.7), (0.7, 0.3), and (0.5, 0.5). The evaluation was performed using three datasets — Gas\_Pipeline, WUSTL\_IIoT and Edge\_IIoT — under three distinct data distribution scenarios: Identical, Dirichlet (non-identical), and No-Label-Skew (non-identical), all based on convergence speed with 100 agents. Among the tested configurations, the sliding window size of $1/10$ with $\alpha = 0.7$ and $\beta = 0.3$ consistently yielded slightly better convergence performance compared to larger window sizes. Consequently, this configuration was selected as the most balanced and effective choice for use throughout this study.

        
        
      

   \begin{table}[ht]
    \centering
    \footnotesize
    \captionsetup{justification=centering} 
    \caption{Sliding-Window Size Selection in DyHFL Based on Convergence Speed with 100 Agents}
    \label{table:slidegas}
    \begin{tabular}{|>{\centering\arraybackslash}p{1.4cm}|
                    >{\centering\arraybackslash}p{1cm}|
                    >{\centering\arraybackslash}p{1.1cm}|
                    >{\centering\arraybackslash}p{0.9cm}|
                    >{\centering\arraybackslash}p{0.9cm}|
                    >{\centering\arraybackslash}p{0.9cm}|}
        \Xhline{1.2pt} 
        \textbf{Dataset} & \textbf{SW Size} & \textbf{$\alpha$ / $\beta$} & \textbf{Dirichlet} & \textbf{No-Label-Skew} & \textbf{Identical} \\ 
        \Xhline{1.2pt} 
        \multirow{9}{*}{Gas\_Pipeline} 
            & \multirow{3}{*}{1/2} & 0.3 / 0.7 & 250  & 58  & 70 \\ \cline{3-6}
            &  & 0.7 / 0.3 & 296  & 58  & 70 \\ \cline{3-6}
            &  & 0.5 / 0.5 & 296  & 56  & 70 \\ \cline{2-6}
            & \multirow{3}{*}{1/5}  & 0.3 / 0.7 & 250  & 55 & 62 \\ \cline{3-6}
            &  & 0.7 / 0.3 & 250  & 60 & 62 \\ \cline{3-6}
            &  & 0.5 / 0.5 & 250  & 78 & 62 \\ \cline{2-6}
            & \multirow{3}{*}{1/10}  & 0.3 / 0.7 & 235 & 57  & 67\\ \cline{3-6}
            &  & 0.7 / 0.3 & 232 & 55  & 67\\ \cline{3-6}
            &  & 0.5 / 0.5 & 250 & 56  & 67\\  
        \Xhline{1.2pt} 
        \multirow{9}{*}{WUSTL\_IIoT}  
            & \multirow{3}{*}{1/2}  & 0.3 / 0.7 & 10 & 9  & 9 \\ \cline{3-6}
            &  & 0.7 / 0.3 & 10 & 8 & 9 \\ \cline{3-6}
            &  & 0.5 / 0.5 & 8 & 7  & 9 \\ \cline{2-6}
            & \multirow{3}{*}{1/5} & 0.3 / 0.7 & 8 & 7 & 7 \\ \cline{3-6}
            &  & 0.7 / 0.3 & 9 & 9 & 7\\ \cline{3-6}
            &  & 0.5 / 0.5 & 10 & 6 & 7\\ \cline{2-6}
            & \multirow{3}{*}{1/10}  & 0.3 / 0.7 & 9  & 9  & 8\\ \cline{3-6}
            &  & 0.7 / 0.3 & 8  & 6  & 8\\ \cline{3-6}
            &  & 0.5 / 0.5 & 8  & 9  & 8 \\ 
        \Xhline{1.2pt} 
            \multirow{9}{*}{Edge\_IIoT}  
            & \multirow{3}{*}{1/2}  & 0.3 / 0.7 & 150 & 25  & 23 \\ \cline{3-6}
            &  & 0.7 / 0.3 & 150 & 25& 23 \\ \cline{3-6}
            &  & 0.5 / 0.5 & 150 & 21 & 24 \\ \cline{2-6}
            & \multirow{3}{*}{1/5} & 0.3 / 0.7 & 125& 23 & 24 \\ \cline{3-6}
            &  & 0.7 / 0.3 & 120 & 20 & 22\\ \cline{3-6}
            &  & 0.5 / 0.5 & 122 & 21 & 22\\ \cline{2-6}
            & \multirow{3}{*}{1/10}  & 0.3 / 0.7 & 100  & 21 & 25\\ \cline{3-6}
            &  & 0.7 / 0.3 & 96  & 20  & 20\\ \cline{3-6}
            &  & 0.5 / 0.5 & 98  & 22 & 22 \\ 
        \Xhline{1.2pt} 
    \end{tabular}
\end{table}

The second step is to determine the appropriate buffer size for achieving appropriate FedBuff results. Buffer sizes of $4$, $10$, and $15$ were tested, representing $25\%$, $50\%$, and $75\%$ of the 20 participant agents. Using the previously defined target accuracy, the best buffer size was determined by examining the convergence speed. The results in Table  \ref{table:fedbuffsizewustl} indicate that FedBuff faces challenges when dealing with buffer sizes of four and ten. These two buffer sizes (i.e., 4 and 10) perform poorly because they do not allow sufficient diversity in updates before aggregation. The situation is particularly problematic in non-identical data settings, where data distribution varies among agents. As an example, in the Dirichlet setting, the model failed to converge with these buffer sizes for both datasets.  When there are not sufficient updates, the model may become overfit or biased toward certain agents, which negatively affects its ability to generalize. Alternatively, large buffer sizes, such as 15, have better results since there are enough updates to aggregate. Therefore, an ideal buffer size for FedBuff is 15 (75\% of the participant agents) because this maintains sufficient diversity resulting in faster convergence and robust model performance.

\begin{table}[ht]
\centering
\footnotesize
\captionsetup{justification=centering} 
\caption{Buffer Size Selection in FedBuff Based on Convergence Speed}
\label{table:fedbuffsizewustl}
\resizebox{\linewidth}{!}{
\begin{tabular}{|c|c|c|c|c|}
\Xhline{1.2pt} 
\textbf{Dataset} & \textbf{SW Size} & \textbf{Identical} & \textbf{Dirichlet} & \textbf{No-Label-Skew} \\ 
\Xhline{1.2pt} 

\multirow{3}{*}{Gas\_Pipeline} & 4  & 80  & not converged  & 284 \\ \cline{2-5}
                               & 10 & 59  & not converged  & 105 \\ \cline{2-5}
                               & 15 & 63  & 400            & 61  \\  
\Xhline{1.2pt} 

\multirow{3}{*}{WUSTL\_IIoT}   & 4  & 87  & not converged  & 62  \\ \cline{2-5}
                               & 10 & 95  & not converged  & 125 \\ \cline{2-5}
                               & 15 & 104 & 590            & 60  \\ 
\Xhline{1.2pt} 

\multirow{3}{*}{\textbf{Edge\_IIoT}} & 4  & 20  & not converged  & 90 \\ \cline{2-5}
                                    & 10 & 15  & not converged  & 85 \\ \cline{2-5}
                                    & 15 & 10  & 500  & 60 \\ 
\Xhline{1.2pt} 
\end{tabular}
}
\end{table}

The last step is to conduct the experiments and assess the results obtained using the defined evaluation metrics.

\subsubsection{\textbf{Agent Selection Fairness}}

                 \begin{figure*}[htbp]
                    \centering
                    \includegraphics[width=7.3in]{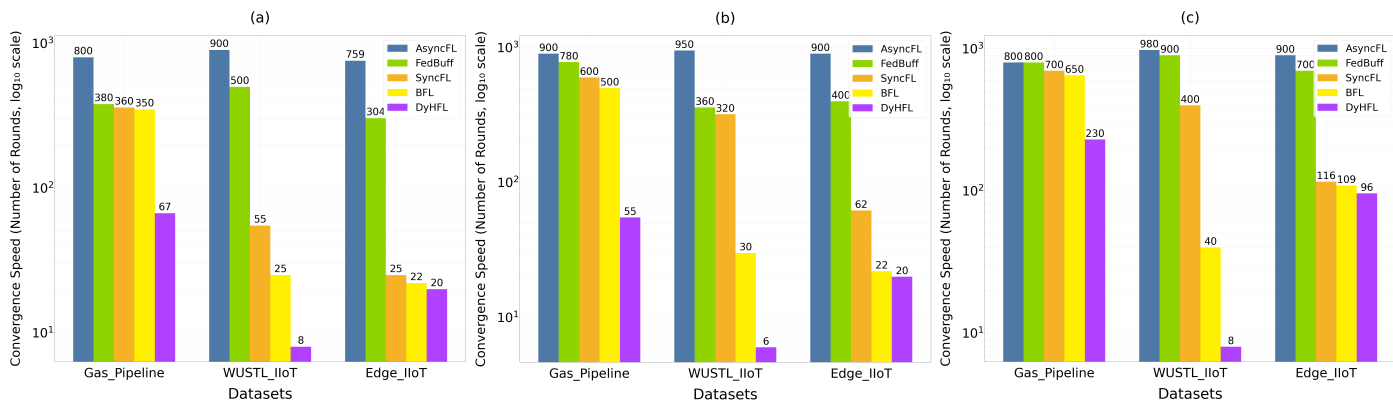}
                    \caption{Convergence Speed Comparison Across FL Methods based on 100 Agents}: (a) Identical Datasets, (b) non-identical datasets with No-Label-Skew distribution, (c) non-identical datasets with Dirichlet distribution. The y-axis indicates the number of FL rounds required to reach the target accuracy, plotted on a logarithmic scale.
                    \label{figure:convgasnonidentical}
                    \vspace{-0.1cm}
                \end{figure*}  
    
    To evaluate the performance of the different FL methods in terms of the defined SRS and FRS metrics, we consider various percentages of straggler agents. Specifically, we analyze scenarios where straggler agents constitute 10\%, 20\%, up to 90\% of the total agent population, across different total agent counts (i.e., 10, 20, ..., 50 agents). Through 20 rounds, the delay times of fast agents were simulated using random integer values between 1 and 5 and those of stragglers were simulated using random integer values between 6 and 10.

                     \begin{table*}[ht]
            \scriptsize
            \caption{Straggler Agent Percentage in BFL}
            \label{table:stragglerold}
            \centering
            \begin{tabular}{|c|cc|cc|cc|cc|cc|}
            \hline
            \multirow{2}{*}{\textbf{Straggler Agents Percentage}} & \multicolumn{2}{c|}{\textbf{10 Agents}} & \multicolumn{2}{c|}{\textbf{20 Agents}} & \multicolumn{2}{c|}{\textbf{30 Agents}} & \multicolumn{2}{c|}{\textbf{40 Agents}} & \multicolumn{2}{c|}{\textbf{50 Agents}} \\ 
            \cline{2-11}
            & \textbf{SRS} & \textbf{FRS} & \textbf{SRS} & \textbf{FRS} & \textbf{SRS} & \textbf{FRS} & \textbf{SRS} & \textbf{FRS} & \textbf{SRS} & \textbf{FRS} \\ \hline
            10\% & 0\% & 66.66\% & 0\% & 72.22\% & 0\% & 85\% & 0\% & 88.87\% & 0\% & 93.33\% \\ \hline
            20\% & 0\% & 100\% & 0\% & 100\% & 0\% & 100\% & 0\% & 100\% & 0\% & 100\% \\ \hline
            30\% & 0\% & 100\% & 16.66\% & 100\% & 22.22\% & 100\% & 25\% & 100\% & 26.66\% & 100\% \\ \hline
            40\% & 25\% & 100\% & 25\% & 100\% & 25\% & 100\% & 31\% & 100\% & 40\% & 100\% \\ \hline
            50\% & 20\% & 100\% & 30\% & 100\% & 33.33\% & 100\% & 35\% & 100\% & 44\% & 100\% \\ \hline
            60\% & 16.66\% & 100\% & 41.66\% & 100\% & 44.44\% & 100\% & 45.83\% & 100\% & 46.66\% & 100\% \\ \hline
            70\% & 28.57\% & 100\% & 42.85\% & 100\% & 47.61\% & 100\% & 57.14\% & 100\% & 62.85\% & 100\% \\ \hline
            80\% & 37.5\% & 100\% & 50\% & 100\% & 54.16\% & 100\% & 59.37\% & 100\% & 70\% & 100\% \\ \hline
            90\% & 44.44\% & 100\% & 50\% & 100\% & 66.66\% & 100\% & 69.44\% & 100\% & 71.11\% & 100\% \\ \hline
            \textbf{Average} & \textbf{19.13\%} & \textbf{96.29\%} & \textbf{26.6\%} & \textbf{96.9\%} & \textbf{30.13\%} & \textbf{98.33\%} & \textbf{33.08\%} & \textbf{98.76\%} & \textbf{37.18\%} & \textbf{99.25\%} \\ \hline

            \end{tabular}
            \normalsize
            \end{table*}

            \begin{table*}[ht]
            \scriptsize
            \caption{Straggler Agent Percentage in DyHFL}
            \label{table:stragglernew}
            \centering
            \begin{tabular}{|c|cc|cc|cc|cc|cc|}
            \hline
            \multirow{2}{*}{\textbf{Straggler Agents Percentage}} & \multicolumn{2}{c|}{\textbf{10 Agents}} & \multicolumn{2}{c|}{\textbf{20 Agents}} & \multicolumn{2}{c|}{\textbf{30 Agents}} & \multicolumn{2}{c|}{\textbf{40 Agents}} & \multicolumn{2}{c|}{\textbf{50 Agents}} \\ 
            \cline{2-11}
            & \textbf{SRS} & \textbf{FRS} & \textbf{SRS} & \textbf{FRS} & \textbf{SRS} & \textbf{FRS} & \textbf{SRS} & \textbf{FRS} & \textbf{SRS} & \textbf{FRS} \\ \hline
            10\% & 0\% & 100\% & 0\% & 100\% & 0\% & 100\% & 0\% & 100\% & 0\% & 100\% \\ \hline
            20\% & 0\% & 100\% & 25\% & 100\% & 33.33\% & 100\% & 25\% & 100\% & 50\% & 100\% \\ \hline
            30\% & 33\% & 100\% & 50\% & 100\% & 55.55\% & 100\% & 50\% & 100\% & 60\% & 100\% \\ \hline
            40\% & 50\% & 100\% & 62.5\% & 100\% & 66.66\% & 100\% & 68.75\% & 100\% & 65\% & 100\% \\ \hline
            50\% & 60\% & 100\% & 70\% & 100\% & 73.33\% & 100\% & 75\% & 100\% & 72\% & 100\% \\ \hline
            60\% & 66.66\% & 100\% & 75\% & 100\% & 77.77\% & 100\% & 79.16\% & 100\% & 76.66\% & 100\% \\ \hline
            70\% & 71.42\% & 100\% & 78.57\% & 100\% & 80.95\% & 100\% & 85.5\% & 100\% & 82.85\% & 100\% \\ \hline
            80\% & 75\% & 100\% & 81.25\% & 100\% & 83.33\% & 100\% & 85\% & 100\% & 74\% & 100\% \\ \hline
            90\% & 77.77\% & 100\% & 77.77\% & 100\% & 77.77\% & 100\% & 77.77\% & 100\% & 77.77\% & 100\% \\ \hline
            \textbf{Average} & \textbf{47.09\%} & \textbf{100\%} & \textbf{53.67\%} & \textbf{100\%} & \textbf{60.61\%} & \textbf{100\%} & \textbf{60.7\%} & \textbf{100\%} & \textbf{62.13\%} & \textbf{100\%} \\ \hline

            \end{tabular}
            \normalsize
            \end{table*}

      According to the results presented in Tables \ref{table:stragglerold} and \ref{table:stragglernew}, DyHFL outperforms the BFL method introduced in \cite{samiracape} by using a sliding window mechanism, making the agent selection process more dynamic and responsive to changes in agent performance over time. This adaptive approach better handles variations in agent performance, especially in environments where agent availability and performance can fluctuate. In terms of SRS, the proposed method achieves an average of 56.44\%, nearly double that achieved by the BFL method (i.e., 29.62\%). This substantial improvement indicates that DyHFL is far more effective in managing stragglers across various agent groups. For example, in Table \ref{table:stragglernew}, a gradual increase in SRS as the straggler agent percentage rises can be observed. For lower percentages of straggler agents, SRS remains at 0\%, but as the percentage of stragglers increases, SRS begins to rise, indicating a more adaptive selection process. In contrast, SRS for the BFL method remains at 0\% for up to 50\% straggler agents, suggesting that its initial round-based selection is less adaptive and might not effectively handle new stragglers appearing in later rounds. Both methods maintain high FRS, but DyHFL does so more effectively across a wider range of agent counts and straggler percentages. It achieves a 100\% rate across all agent configurations, while the BFL method averages 97.91\%. Although this difference might seem small, DyHFL guarantees FRS consistently, making it more reliable. This suggests that the new method's dynamic selection process helps consistently in identifying and leveraging fast agents for efficient training.

    In both SRS and FRS, DyHFL outperforms the BFL method by incorporating a sliding window mechanism that adapts to changes in the agent performance. In environments with varying agent performance, this approach allows for better management of stragglers and more consistent utilization of fast agents.

        \begin{table*}[ht]
        \centering
        \footnotesize
        \caption{Two-Tailed t-Test p-Values Comparing DyHFL with Other FL Baselines on Communication Cost}
        \label{table:comcostpval}
        \begin{tabular}{|c|c|c|c|c|c|c|}
        \hline
        \textbf{Dataset} & \textbf{Agents} & \textbf{BFL} & \textbf{Fedbuff} & \textbf{ASR\_Fed} & \textbf{Async} & \textbf{Sync} \\
        \hline
        \multirow{4}{*}{Gas\_Pipeline}
        & 20  & $1.05\times10^{-5}$ & $8.43\times10^{-5}$ & $1.38\times10^{-5}$ & $2.88\times10^{-9}$ & $4.97\times10^{-6}$ \\
        & 40  & $2.81\times10^{-6}$ & $7.15\times10^{-7}$ & $8.81\times10^{-10}$ & $3.47\times10^{-11}$ & $5.24\times10^{-8}$ \\
        & 80  & $9.49\times10^{-5}$ & $9.35\times10^{-5}$ & $3.30\times10^{-7}$ & $9.35\times10^{-11}$ & $2.90\times10^{-7}$ \\
        & 100 & $1.25\times10^{-2}$ & $3.18\times10^{-3}$ & $2.17\times10^{-3}$ & $7.27\times10^{-9}$  & $2.58\times10^{-5}$ \\
        \hline
        \multirow{4}{*}{WUSTL\_IIoT}
        & 20  & $7.11\times10^{-4}$  & $6.92\times10^{-5}$ & $1.17\times10^{-5}$ & $1.93\times10^{-9}$ & $3.59\times10^{-6}$ \\
        & 40  & $1.81\times10^{-3}$ & $2.99\times10^{-4}$ & $1.25\times10^{-4}$ & $1.41\times10^{-8}$ & $2.16\times10^{-5}$ \\
        & 80  & $3.09\times10^{-2}$ & $8.33\times10^{-4}$ & $4.23\times10^{-5}$ & $1.14\times10^{-9}$ & $4.55\times10^{-6}$ \\
        & 100 & $4.55\times10^{-2}$ & $6.46\times10^{-6}$ & $2.17\times10^{-7}$ & $2.49\times10^{-11}$ & $8.10\times10^{-8}$ \\
        \hline
        \multirow{4}{*}{Edge\_IIoT}
        & 20  & $3.11\times10^{-2}$ & $2.13\times10^{-7}$ & $3.47\times10^{-9}$ & $1.62\times10^{-10}$ & $1.72\times10^{-7}$ \\
        & 40  & $1.04\times10^{-4}$ & $3.04\times10^{-8}$ & $4.74\times10^{-7}$ & $1.27\times10^{-11}$ & $1.41\times10^{-8}$ \\
        & 80  & $8.40\times10^{-5}$ & $2.01\times10^{-9}$ & $1.44\times10^{-15}$ & $6.09\times10^{-10}$ & $1.25\times10^{-9}$ \\
        & 100 & $3.02\times10^{-2}$ & $2.80\times10^{-8}$ & $2.54\times10^{-8}$ & $1.82\times10^{-11}$ & $2.27\times10^{-8}$ \\
        \hline
        \end{tabular}
        \end{table*}


\begin{table*}[ht]
\centering
\footnotesize
\caption{Two-Tailed t-Test p-Values Comparing DyHFL with Other FL Baselines on Convergence Speed}
\label{table:pvalues_convergence}
\begin{tabular}{|c|c|c|c|c|c|}
    \Xhline{1.2pt}
    \textbf{Dataset} & \textbf{Data Type} & \textbf{Async} & \textbf{Sync} & \textbf{Fedbuff} & \textbf{BFL} \\
    \Xhline{1.2pt}
    
    \multirow{3}{*}{\textbf{Gas\_Pipeline}} 
        & Identical & \textbf{$3.49\times10^{-9}$} & \textbf{$9.28\times10^{-6}$} & \textbf{$2.36\times10^{-10}$} & \textbf{$6.96\times10^{-10}$} \\

        & No-Label-Skew & \textbf{$2.49\times10^{-10}$} & \textbf{$3.084\times 10^{-4}$} & \textbf{$2.28\times10^{-7}$} & \textbf{$2.55\times10^{-8}$} \\
 
        & Dirichlet & \textbf{$3.19\times10^{-9}$} & \textbf{$8.68\times10^{-7}$} & \textbf{$4.64\times10^{-7}$} & \textbf{$3.92\times10^{-7}$} \\

    \Xhline{1.2pt}

    \multirow{3}{*}{\textbf{WUSTL\_IIoT}} 
        & Identical  &  $7.46\times10^{-11}$ & $1.28\times10^{-4}$ & $1.05\times10^{-7}$ & $9.00\times10^{-9}$ \\
        &    No-Label-Skew  &$1.62\times10^{-7}$ & $5.27\times10^{-6}$ & $3.19\times10^{-7}$ & $9.51\times10^{-6}$ \\
        & Dirichlet &  $3.75\times10^{-11}$ & $2.50\times10^{-6}$ & $2.45\times10^{-7}$ & $1.31\times10^{-6}$   \\
    \Xhline{1.2pt}

    \multirow{3}{*}{\textbf{Edge\_IIoT}} 
        & Identical  & $7.103\times 10^{-7}$ & $7.18\times 10^{-3}$ & $7.224\times 10^{-7}$ & $1.747\times 10^{-2}$ \\
        & No-Label-Skew  &  $7.798\times 10^{-11}$ & $2.620\times 10^{-5}$ & $9.706\times 10^{-10}$ & $1.038\times 10^{-2}$ \\
        & Dirichlet  & $5.824\times 10^{-8}$ & $8.064\times 10^{-4}$ & $1.374\times 10^{-7}$ & $8.109\times 10^{-3}$ \\
        
    \Xhline{1.2pt}
\end{tabular}
\end{table*}

\subsubsection{\textbf{Convergence Speed}}

    Fig. \ref{figure:convgasnonidentical} illustrates the convergence speed for different models on both identical and non-identical datasets for 100 agents.

  According to Fig. \ref{figure:convgasnonidentical}, \textbf{DyHFL}  consistently outperforms the other five methods —AsyncFL, FedBuff, SyncFL, ASR\_Fed, and BFL— on both identically and non-identically distributed datasets, demonstrating its robustness to heterogeneous data. This robustness is due to its balanced approach of selecting both fast and slow agents, ensuring effective learning and balanced participation. For example, in the Gas\_Pipeline dataset, DyHFL converges 11.9 times faster than AsyncFL, 5.6 times faster than FedBuff, 5.3 times faster than SyncFL, and 5.2 times faster than BFL on identical datasets. The performance advantage becomes more pronounced on non-identical datasets, with DyHFL converging 16.3 times faster than AsyncFL, 14.18 times faster than FedBuff,10.9 times faster than SyncFL, and 9 times faster than BFL on no-label-skew data, and 3.4, 3.4, 3, and 2.8 times faster on Dirichlet non-identical data, respectively. Similarly, with the WUSTL\_IIoT dataset, DyHFL surpasses AsyncFL by 112 times, FedBuff by 62.5 times, 6.8 times faster than SyncFL, and BFL by 3.2 times on identical datasets. The advantages are even greater for non-identical data, with DyHFL converging 158 times faster than AsyncFL, 60 times faster than FedBuff, 53 times faster than SyncFL, and 5 times faster than BFL on no-label-skew data, and 122, 112, 50, and 5 times faster on Dirichlet non-identical data, respectively. Regarding Edge\_IIoT dataset, DyHFL surpasses AsyncFL by 37.95 times, FedBuff by 15.2 times, 1.25 times faster than SyncFL, and BFL by 1.1 times on identical datasets. The advantages are even greater for non-identical data, with DyHFL converging 45 times faster than AsyncFL, 20 times faster than FedBuff, 3.1 times faster than SyncFL, and 1.1 times faster than BFL on no-label-skew data, and 9.37, 7.2, 1.20, and 1.13 times faster on Dirichlet non-identical data, respectively.
  
  This clearly demonstrates the DyHFL's efficiency and adaptability across diverse scenarios.

   The differences in  \textbf{DyHFL} convergence results between the three datasets can be attributed to their intrinsic properties. The Gas\_Pipeline dataset is smaller in overall size. As a result, under No-label-skew conditions, due to the limited number of samples, agents receive very homogeneous data, making learning more difficult and convergence slower. Under the Dirichlet distribution, the diversity of the data mitigates some of these effects, leading to relatively better performance. On the other hand, the WUSTL\_IIoT and Edge\_IIoT datasets have a much larger volume of data. Consequently, even under challenging scenarios such as No-label-skew, agents still receive sufficiently diverse and abundant data, resulting in more stable model performance and less variation between different data distributions.

      \textbf{AsyncFL} shows the slowest convergence overall, requiring the most rounds, particularly for non-identical datasets. Its asynchronous nature causes each agent to update independently, which leads to delays and communication bottlenecks. These findings suggest that methods relying on unoptimized communication between the server and agents are less efficient, especially when handling non-identical data. \textbf{FedBuff} requires significantly more rounds to converge on non-identical datasets than on identical ones, indicating a struggle to manage diverse data distributions effectively. The buffer mechanism used in the aggregator server contributes to inefficiencies and delays, as faster agents with less training data tend to fill the buffer quickly, causing imbalances and biases in the training process. In comparison to FedBuff and AsyncFL, \textbf{BFL} performs better on both identical and non-identical datasets, as it employs an agent selection method to mitigate the straggler effect. However, since agent selection is only applied in the first round and lacks a dynamic approach within the FL framework, it is somewhat inefficient and less effective compared to DyHFL.

     The \textbf{ASR\_Fed} approaches failed to converge effectively in the experiments, particularly when the number of agents increased to 100. ASR\_Fed, while designed to mitigate the straggler problem by prioritizing fast agents (buffer agents) and incorporating updates from slow agents (circumvent agents) only after a computed threshold, still suffers in high-agent scenarios. When the number of agents scales up, the threshold for including slower agents in ASR\_Fed increases accordingly, causing many agents to be excluded from early aggregation rounds. This leads to insufficient global representation during training and consequently slows down or prevents convergence altogether.

    Table~\ref{table:pvalues_convergence} presents the two-tailed t-test p-values from the comparison of DyHFL with other FL baselines (Async, Sync, Fedbuff, and BFL) in terms of convergence speed, evaluated across both identical and non-identical datasets.
    
    The \textbf{two-tailed t-test p-value} is a statistical measure used to determine whether the observed differences between two methods are statistically significant or could have occurred by random chance. A p-value less than 0.05 indicates that the performance difference is statistically significant at the 95\% confidence level. 
    
    The results in Table~\ref{table:pvalues_convergence} show that all p-values are far below the 0.05 threshold, often reaching the order of $10^{-7}$ to $10^{-20}$. This confirms that DyHFL’s improvements in convergence speed are not due to random variation but are statistically reliable. 
 
     Based on the consistent performance of DyHFL across all settings demonstrates that DyHFL significantly outperforms other baselines regardless of dataset type or data heterogeneity. This validates its robustness and generalizability in real-world IIoT scenarios, driven by its dynamic approach to balancing participation between fast and slow agents.

\subsubsection{\textbf{Communication Cost}}   Table \ref{table:parameters} shows the settings that have been considered for this calculation. 
                {\begin{table}[h!]
                \caption{Experimental Parameters}
                \label{table:parameters}
                \centering
                \begin{tabular}{|l|l|}
                \hline
                \textbf{Parameter} & \textbf{Value} \\ \hline
                Rounds & 10 \\ \hline
                Local epochs & 10 \\ \hline
                Number of agents & 20,40,80,100\\ \hline
                Sliding window & 1/10 of rounds \\ \hline
                $\alpha$ and $\beta$ & 0.7 and 0.3 \\ \hline
                Buffer size for Feddbuff & 75\% of total agents \\ \hline
                Gas\_Pipeline dataset Model size& 0.009 MB \\ \hline
                WUSTL\_IIoT dataset Model size & 0.002 MB \\ \hline
                Edge\_IIoT dataset Model size & 0.014 MB \\ \hline
                \end{tabular}
                \end{table}


            According to Fig. \ref{figure:comcosttt}, and compared to other algorithms, the \textbf{DyHFL} reduces communication costs by dividing the total communication cost into preliminary and subsequent rounds. The result of this optimization is lower communication costs, especially when there are a lot of agents.

            The \textbf{BFL} method also reduces communication costs by dividing total communication costs between the first and subsequent rounds. However, the BFL approach is not as effective as DyHFL, which results in higher communication costs. \textbf{Fedbuff} reduces communication costs by buffering and transmitting updates from a fraction of model updates (75\% of total agents), which also minimizes model transmissions. While Fedbuff’s buffering reduces communication overhead, it may slow convergence due to potential biases from agents with less data or lower accuracy. In contrast, \textbf{SyncFL} transmits updates from all agents in each round, resulting in higher communication costs. SyncFL's comprehensive aggregation captures diverse data patterns, leading to faster convergence despite higher costs.

                                                   \begin{figure}[htbp]
                                        \centerline{\includegraphics[width=3.5in]{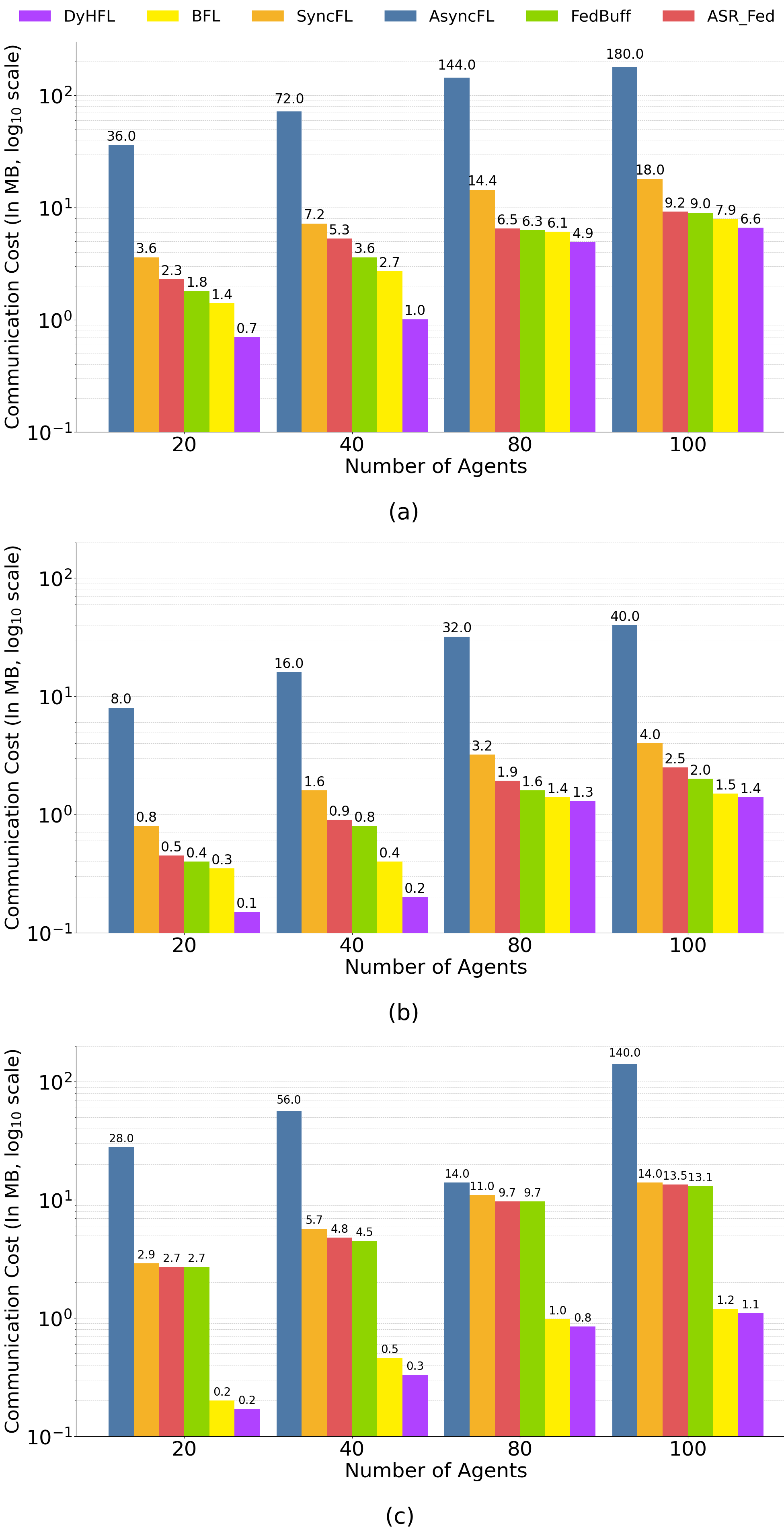}}
                                        \captionsetup{justification=centering}
                                        \caption{Comparison of communication cost (exchanged message size in megabytes, presented on a logarithmic scale)} across agents in FL methods: (a) Gas\_Pipeline Dataset, (b) WUSTL\_IIoT Dataset, and (c) Edge\_IIoT Dataset.
                                        \label{figure:comcosttt}
                                        \vspace{-0.1cm}
                                    \end{figure}

            \textbf{AsyncFL} has the worse results as a consequence of frequent communications with the server. AsyncFL's frequent communications significantly increase its communication costs, making it the least efficient of the other methods.

            The communication cost of \textbf{ASR\_Fed} is lower compared to \textbf{AsyncFL} and \textbf{SyncFL}, as it prioritizes fast agents and delays the participation of slow agents. This selective participation reduces the frequency and volume of communication, especially in the early rounds, leading to a better communication cost.

            Table~\ref{table:comcostpval} presents the two-tailed t-test p-values comparing DyHFL against other FL baselines in terms of communication cost across three datasets and varying agent numbers. In all cases, the p-values are well below the standard significance threshold} ($p < 0.05$), confirming that the observed differences are statistically significant. Notably, DyHFL consistently achieves lower communication costs, with extremely small p-values (as low as $10^{-11}$) when compared to AsyncFL and ASR\_Fed, highlighting a substantial performance gap. The consistent statistical significance across datasets and agent scales reinforces the robustness of DyHFL’s communication efficiency advantage.

\subsubsection{\textbf{Model Performance}}    

\begin{figure*}[htbp]
                                \centering
                               
                                \includegraphics[width=\textwidth]{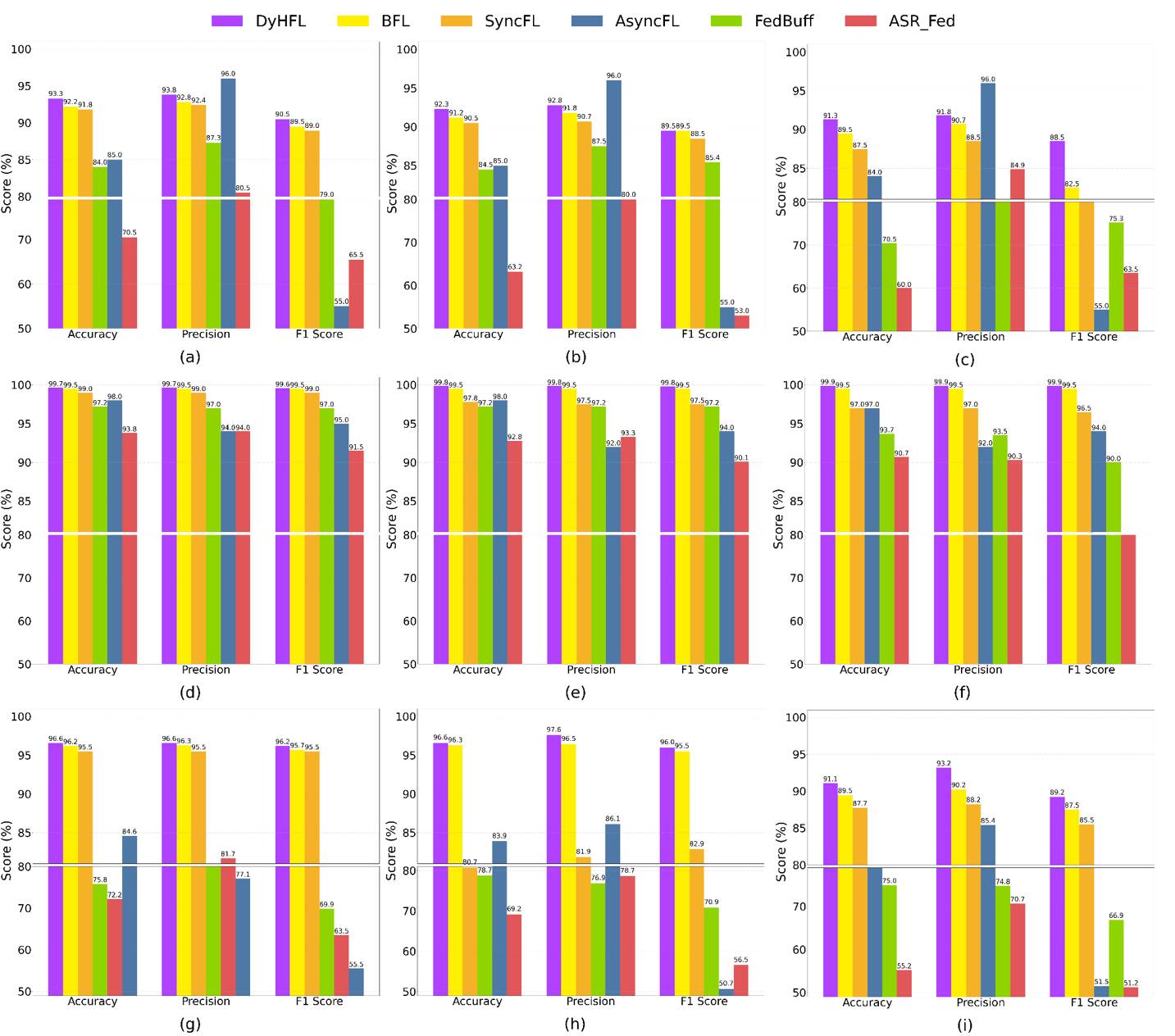}
                                
                                \captionsetup{justification=centering}
                                \caption{Model performance comparison of six FL algorithms (SyncFL, AsyncFL, FedBuff, BFL, ASR\_Fed, and DyHFL) across different datasets and conditions: $(a)$–$(c)$ Gas\_Pipeline dataset under identical, non-identical with No-Label-Skew distribution, and non-identical with Dirichlet distribution settings; $(d)$–$(f)$ WUSTL\_IIoT dataset under the same three settings; and $(g)$–$(i)$ Edge\_IIoT dataset under the same three settings.}

                                \label{figure:acccomparisoncombined}
                                \vspace{-0.1cm}
                            \end{figure*}

    Fig. \ref{figure:acccomparisoncombined} compares six FL algorithms - SyncFL, AsyncFL, FedBuff, ASR\_Fed, BFL, and DyHFL - on the mentioned datasets with 100 agents and 10 rounds, demonstrating their performance through key metrics: Accuracy, Precision, and F1 Score. In the image, each subfigure corresponds to one of three conditions (identical, no-label-skew non-identical, Dirichlet non-identical), and reveals key insights about FL baseline effectiveness in handling identical and non-identical data distributions.

        DyHFL consistently shows the best performance across all datasets and conditions, scoring high across all metrics. Its success lies in its balanced selection of agents, which ensures a fair representation of both fast and slow agents. With this strategy, DyHFL is able to handle diverse and imbalanced data distributions more effectively than other algorithms.

        BFL achieves similar results to DyHFL due to its approach in selecting agents that ensures fairness between them. However, BFL's results are not as optimal as DyHFL's because BFL's agent selection method lacks the dynamic adaptability of DyHFL, which makes it less effective.

        SyncFL also performs well, with results similar to DyHFL's. This is largely due to its synchronous communication method, which ensures uniform model updates across all agents. SyncFL does not include the fairness adjustments that give DyHFL a slight advantage, especially when handling diverse data.
        
        AsyncFL generally performs worse, especially in scenarios with imbalanced datasets, like the Gas\_Pipeline dataset, or with highly non-identical data distributions. Its asynchronous approach struggles to handle the challenges posed by such data characteristics, resulting in lower F1 scores. While AsyncFL achieves high precision on the Gas\_Pipeline dataset, this is due to the dataset's imbalance, which can lead to misleading results.
        
        FedBuff maintains a competitive position. However, its buffering mechanism poses challenges, especially when dealing with non-identical datasets. In FedBuff, the buffering strategy favors data from faster agents, which can lead to a poor generalization and training bias.

        The model performance of \textbf{ASR\_Fed} is, on average, consistently the lowest across all scenarios and datasets, as shown in Figures~(a)–(i). This underperformance can be attributed to its selective aggregation strategy, where only fast agents participate in the early rounds. While this reduces communication cost and mitigates stragglers, it also delays the inclusion of slower agents (circumvent agents), leading to limited representation of the overall data distribution. Consequently, the model lacks diversity in the aggregated updates, which negatively impacts generalization and ultimately results in poorer model performance compared to the other approaches.
        
         The results indicate that DyHFL is more effective for FL tasks with diverse and imbalanced data in a heterogeneous environment. Despite the similar convergence speed and model performance of DyHFL and BFL, DyHFL excels in other important aspects, such as agent selection fairness and communication cost, making it more suitable for real-world applications. This shows that DyHFL is more stable in the presence of straggler agents while preserving performance.

\subsubsection {\textbf{Component Ablation Analysis}}

To rigorously evaluate the individual contributions of HE and dynamic agent selection to the overall performance of the proposed DyHFL framework, an ablation study was conducted in which each component —either HE or dynamic agent selection—was selectively removed. 

In this analysis, the variants with and without encryption are denoted as \textit{DyHFL Paillier} and \textit{DyHFL Plain}, respectively, while the absence of dynamic selection reverts the framework to the baseline synchronous aggregation, referred to as \textit{SyncFL Paillier} and \textit{SyncFL Plain}. By disabling one component (HE and dynamic agent selection ) at a time and monitoring changes in model performance, convergence speed, and communication cost, the study provides a clear quantitative assessment of the role played by each component. Tables~\ref{table:ablationiden}, \ref{table:ablationNolablel}, and \ref{table:AblationDrichlet} present the ablation results across all settings.

\begin{table*}[ht]
\centering
\footnotesize
\setlength{\tabcolsep}{3pt}
\caption{\small Component Ablation Analysis on Identical Datasets}
\label{table:ablationiden}
\begin{tabular}{|
>{\centering\arraybackslash}p{1.6cm}|
>{\centering\arraybackslash}p{1.9cm}|
>{\centering\arraybackslash}p{1.8cm}|
>{\centering\arraybackslash}p{1.9cm}|
>{\centering\arraybackslash}p{1.2cm}|
>{\centering\arraybackslash}p{1.2cm}|
>{\centering\arraybackslash}p{1.2cm}|}
\hline
\textbf{Dataset} & \textbf{Algorithms} & \textbf{Convergence Speed} & \textbf{Communication Cost} & \multicolumn{3}{c|}{\textbf{Model Performance}} \\
\cline{5-7}
 & & & & \textbf{Accuracy} & \textbf{Precision} & \textbf{F1-score}\\
\hline
\multirow{4}{*}{Gas\_Pipeline}
& DyHFL Pailliar & 67  & 0.66 & 93.3 & 93.8 & 90.5 \\
& DyHFL Plain    & 66  & 0.66 & 94.3 & 94.8 & 91.5 \\
& Sync Pailliar  & 360 & 3.60 & 91.8 & 92.4 & 89.0 \\
& Sync Plain     & 360 & 3.60 & 91.8 & 92.4 & 89.0 \\
\hline
\multirow{4}{*}{WUSTL\_IIoT}
& DyHFL Pailliar & 8   & 0.15 & 99.7 & 99.7 & 99.6 \\
& DyHFL Plain    & 8   & 0.15 & 99.7 & 99.7 & 99.6 \\
& Sync Pailliar  & 55  & 0.80 & 99.0 & 99.0 & 99.0 \\
& Sync Plain     & 54  & 0.80 & 98.4 & 99.2 & 98.2 \\
\hline
\multirow{4}{*}{Edge\_IIoT}
& DyHFL Pailliar & 20  & 0.17 & 96.6 & 96.6 & 96.2 \\
& DyHFL Plain    & 21  & 0.17 & 94.6 & 95.6 & 95.2 \\
& Sync Pailliar  & 25  & 2.90 & 95.5 & 95.5 & 95.5 \\
& Sync Plain     & 25  & 2.90 & 95.5 & 95.5 & 95.5 \\
\hline
\end{tabular}
\end{table*}

\begin{table*}[ht]
\centering
\footnotesize
\setlength{\tabcolsep}{3pt}
\caption{\small  Component Ablation Analysis on No-label-skew Non-Identical Datasets}
\label{table:ablationNolablel}
\begin{tabular}{|
>{\centering\arraybackslash}p{1.6cm}|
>{\centering\arraybackslash}p{1.9cm}|
>{\centering\arraybackslash}p{1.8cm}|
>{\centering\arraybackslash}p{1.9cm}|
>{\centering\arraybackslash}p{1.2cm}|
>{\centering\arraybackslash}p{1.4cm}|
>{\centering\arraybackslash}p{1.2cm}|}
\hline
\textbf{Dataset} & \textbf{Algorithms} & \textbf{Convergence Speed} & \textbf{Communication Cost} & \multicolumn{3}{c|}{\textbf{Model Performance}} \\
\cline{5-7}
 & & & & \textbf{Accuracy} & \textbf{Precision} & \textbf{F1-score} \\
\hline
\multirow{4}{*}{Gas\_Pipeline}
& DyHFL Pailliar & 55  & 0.66 & 92.3 & 92.8 & 89.5 \\
& DyHFL Plain    & 53  & 0.66 & 92.7 & 93.2 & 90.5 \\
& Sync Pailliar  & 650 & 3.60 & 90.5 & 90.7 & 88.5 \\
& Sync Plain     & 648 & 3.60 & 91.3 & 92.2 & 89.7 \\
\hline
\multirow{4}{*}{WUSTL\_IIoT}
& DyHFL Pailliar & 6   & 0.15 & 99.8 & 99.8 & 99.8 \\
& DyHFL Plain    & 6   & 0.15 & 99.8 & 99.8 & 99.8 \\
& Sync Pailliar  & 320 & 0.80 & 97.8 & 97.5 & 97.5 \\
& Sync Plain     & 324 & 0.80 & 96.3 & 93.2 & 95.4 \\
\hline
\multirow{4}{*}{Edge\_IIoT}
& DyHFL Pailliar & 20  & 0.17 & 96.6 & 97.6 & 96.0 \\
& DyHFL Plain    & 20  & 0.17 & 96.6 & 97.6 & 96.0 \\
& Sync Pailliar  & 62  & 0.29 & 80.7 & 81.9 & 82.9 \\
& Sync Plain     & 61  & 0.29 & 80.7 & 81.9 & 82.9 \\
\hline
\end{tabular}
\end{table*}

\begin{table*}[ht]
\centering
\footnotesize
\setlength{\tabcolsep}{3pt}
\caption{\small Component Ablation Analysis on Dirichlet Non-Identical Datasets}
\label{table:AblationDrichlet}
\begin{tabular}{|
>{\centering\arraybackslash}p{1.6cm}|
>{\centering\arraybackslash}p{1.9cm}|
>{\centering\arraybackslash}p{1.8cm}|
>{\centering\arraybackslash}p{1.9cm}|
>{\centering\arraybackslash}p{1.2cm}|
>{\centering\arraybackslash}p{1.4cm}|
>{\centering\arraybackslash}p{1.2cm}|}
\hline
\textbf{Dataset} & \textbf{Algorithms} & \textbf{Convergence Speed} & \textbf{Communication Cost} & \multicolumn{3}{c|}{\textbf{Model Performance}} \\
\cline{5-7}
 & & & & \textbf{Accuracy} & \textbf{Precision} & \textbf{F1-score} \\
\hline
\multirow{4}{*}{Gas\_Pipeline}
& DyHFL Pailliar & 230 & 0.66 & 91.3 & 91.8 & 88.5 \\
& DyHFL Plain    & 229 & 0.66 & 91.7 & 92.1 & 88.7 \\
& Sync Pailliar  & 700 & 3.60 & 87.5 & 88.5 & 80.8 \\
& Sync Plain     & 703 & 3.60 & 85.2 & 85.3 & 76.4 \\
\hline
\multirow{4}{*}{WUSTL\_IIoT}
& DyHFL Pailliar & 8   & 0.15 & 99.9 & 99.9 & 99.9 \\
& DyHFL Plain    & 8   & 0.15 & 99.9 & 99.9 & 99.9 \\
& Sync Pailliar  & 400 & 0.80 & 99.5 & 99.5 & 99.5 \\
& Sync Plain     & 400 & 0.80 & 99.4 & 99.2 & 99.6 \\
\hline
\multirow{4}{*}{Edge\_IIoT}
& DyHFL Pailliar & 96  & 0.17 & 91.2 & 93.2 & 89.2 \\
& DyHFL Plain    & 93  & 0.17 & 92.7 & 94.8 & 91.3 \\
& Sync Pailliar  & 116 & 0.29 & 87.7 & 88.2 & 85.5 \\
& Sync Plain     & 114 & 0.29 & 88.4 & 88.9 & 86.3 \\
\hline
\end{tabular}
\end{table*}

Across all datasets and scenarios, DyHFL—both in encrypted and plaintext modes—achieves substantially faster convergence speeds compared to the synchronous FL (SyncFL) baseline. For example, in the no-label-skew WUSTL-IIoT dataset, DyHFL converges in only 6 rounds, whereas SyncFL requires over 320 rounds. This significant improvement demonstrates the effectiveness of the proposed dynamic agent selection with buffering in mitigating the straggler effect that typically slows down synchronous aggregation. By adaptively selecting a subset of agents within a sliding window and aggregating updates without waiting for all agents to complete their local training, DyHFL reduces delays caused by slower participants. This strategy not only accelerates convergence—often by up to an order of magnitude in heterogeneous data scenarios—but also directly lowers the overall communication cost, as the total cost is proportional to the total number of agents. The reduction in agents due to the DyHFL agent selection mechanism results in more efficient use of bandwidth. Moreover, DyHFL consistently maintains high accuracy, precision, and F1-score across all data distributions, outperforming SyncFL in both predictive quality and training efficiency. The dynamic selection mechanism further preserves participation diversity across rounds, which is crucial for sustaining model generalization under non-identical data distributions. Together, these results confirm that dynamic agent selection is the primary driver behind DyHFL’s superior performance, enabling faster, more communication-efficient, and more accurate FL without compromising fairness or robustness.

Incorporating Paillier-based HE into DyHFL enables secure aggregation without exposing individual agent updates, providing strong privacy guarantees. The encryption process has a negligible impact on convergence speed or model performance because the Paillier scheme preserves exact numerical values during aggregation, avoiding quantization or approximation errors. Although HE theoretically increases per-round communication size due to ciphertext expansion, in this study the communication cost is computed using a formula based on model size ($M$), number of agents ($N$), total rounds ($T$), and buffer parameters, with $M$ fixed for both encrypted and plaintext models. This approach isolates the algorithmic impact of HE and dynamic selection, resulting in identical reported communication costs for Paillier and Plain variants. In real deployments that account for the true ciphertext size, Paillier would introduce a modest increase in per-round cost; however, DyHFL’s buffering mechanism would still offset this overhead by significantly reducing the total number of rounds, thereby preserving overall communication efficiency while ensuring data privacy.

Overall, the ablation results clearly demonstrate that dynamic agent selection is the dominant factor in enhancing efficiency, while HE strengthens privacy without sacrificing predictive performance or computational efficiency.

\section {\textbf{Discussion and Future Directions}}\label{disc}

Recent advances in privacy-preserving FL motivate positioning DyHFL against emerging alternatives. One such method is Federated Transfer Learning (FTL), which allows collaboration among agents with heterogeneous feature or label spaces. While FTL is well-suited for cross-domain FL scenarios, it often lacks strong aggregation privacy guarantees and relies on pre-trained representations. In contrast, DyHFL assumes aligned input spaces but offers enhanced privacy through HE and ensures fairness through dynamic agent selection based on real-time metrics. This makes DyHFL particularly advantageous in IIoT environments where devices vary in capabilities, but privacy, latency, and coordination remain critical.

In the current design, DyHFL incorporates HE-based updates, threshold evaluation, and buffer management, which introduce additional processing time as well as memory and energy demands that may exceed the capabilities of embedded controllers and low-power sensors. The computational cost of encryption and decryption, combined with ciphertext expansion, also increases bandwidth usage and can place additional load on low-rate industrial links. Furthermore, as with most HE-based FL systems, DyHFL employs a trusted third party (TTP) for key generation and distribution, introducing a centralized element that may not align with the requirements of fully decentralized or adversarial IIoT environments. To mitigate this, several future research directions are proposed:

\begin{itemize}
\item Employ quantized or selectively applied encryption to reduce computational complexity while preserving confidentiality.

\item Adaptive participation mechanisms in which severely resource-constrained devices participate at reduced frequencies, selectively transmit critical model parameters, or employ model update compression techniques to minimize communication and computation overhead.
\item Combine the strong aggregation privacy guarantees of DyHFL with the cross-domain adaptability of FTL to address both statistical and feature-space heterogeneity in large-scale IIoT deployments.

\item Integration of \emph{Distributed Key Generation} (DKG) offers a promising solution to eliminating the single point of failure inherent in centralized key management. DKG allows agents to collaboratively generate cryptographic keys without reliance on a trusted third party. Incorporating DKG would enhance resilience, support dynamic key rotation, and improve security in decentralized or adversarial environments.

\end{itemize}

DyHFL achieves strong aggregation privacy and coordinated convergence in IIoT environments through the integration of HE with dynamic agent selection. The above mitigations form a practical roadmap to broaden DyHFL’s applicability in real-world IIoT.

 \section{Conclusion}\label{conc}
\vspace{-0.06cm}

In this paper, we introduced a new dynamic FL framework designed to detect anomalies, such as cyber threats, in industrial CPSs. The proposed DyHFL framework incorporates a secure communication protocol based on HE to protect model parameters from model inversion attacks. Additionally, an innovative agent selection strategy was developed. It effectively balances the performance of fast and slow agents in heterogeneous environments, minimizing straggler effects and reducing communication bottlenecks. Extensive experiments, conducted using two real-world industrial CPS datasets, demonstrated that DyHFL does not only achieve superior prediction accuracy but also converges more quickly compared to existing FL approaches, proving its effectiveness and efficiency in practical applications. 

\section*{Acknowledgment}
\vspace{-0.06cm}

This work is partly conducted at ICTFICIAL Oy, Finland. It is supported in part by the European Union’s Horizon Europe research and innovation program HORIZON-JU-SNS-2022 under the RIGOUROUS project (Grant No. 101095933), and the 6G-Path project under Grant No. 101139172. The paper reflects only the authors’ views, and the European Commission bears no responsibility for any utilization of the information contained herein.


\bibliographystyle{IEEEtran}
\bibliography{bib}

\end{document}